\documentclass[12pt]{iopart}
\usepackage{lineno,hyperref}
\usepackage{indentfirst}
\usepackage{array}
\usepackage{graphicx,setspace}
\usepackage{amssymb}
\usepackage{epstopdf}
\usepackage{float}
\usepackage{subfigure}

\begin{document}

\title[]{Target Search of a Protein on DNA in the Presence of Position-dependent Bias}

\author{Xi Chen$^{1,2}$, Xiujun Cheng$^3$, Yanmei Kang$^1$, Jinqiao Duan$^{2,*}$}

\address{$^1$ School of Mathematics and Statistics,Xi'an Jiaotong University, Xi'an, 710049, China}
\address{$^2$ Department of Applied Mathematics, Illinois Institute of Technology, Chicago, IL 60616, USA}
\address{$^3$Center for Mathematical Science $\&$School of Mathematics and Statistics, Huazhong University of Science and Technology, Wuhan 430074, China}
\ead{duan@iit.edu}
\vspace{10pt}
\begin{indented}
\item[]September 2018
\end{indented}

\begin{abstract}
We study the target searching on the DNA for proteins in the presence of non-constant drift and non-Gaussian $\alpha$-stable L\'evy fluctuations. The target searching is realized by the facilitated diffusion process. The existing works are about this problem in the case of constant drift. Starting from a non-local Fokker-Planck equation with a ``sink'' term, we obtain the possibility density function for the protein occurring at position $x$ on time $t$. Based on this, we further compute the survival probability and the first arrival density in order to quantify the searching mechanisms. The numerical results show that in the linear drift case, there is an optimal $\alpha$ index for the search to be most likely successful (searching reliability reaches its maximum). This optimal $\alpha$ index depends on initial position-target separation. It is also found that the diffusion intensity plays a positive role in improving the searching success. The nonlinear double-well drift could drive the protein to reach the target with a larger possibility than the linear drag at initial time period, but viewing at the long time evolution, the linear drift is more beneficial for target searching success. In contrast to the linear drift case, the search reliability and efficiency with nonlinear drift have a monotonic relationship with the $\alpha$ index, that is, the smaller the $\alpha$ index is, the more possibly a protein finds its target.
\end{abstract}
\vspace{2pc}
\noindent{\it Keywords\ } {DNA-protein binding, non-local Fokker-Planck equation, L\'evy flights, linear and nonlinear biases, non-local diffusion}
%
%
%
%

\section{Introduction}
The biological functions are essentially realized by the proteins recognising and binding the specific sites on DNA \cite {Riggs1970,Berg1981,Berg1985,Lodish2000,Ptashne2001,Elf2007,Halford2009,Bressloff2013,Bressloff,Marko2015,Niranjani2016}. Familiar examples occur at the restriction enzymes in bacteria, chromatin-remodeling proteins in eukarya, and transcription factors across all domain of life \cite{Lodish2000,Ptashne2001,Bressloff,Marko2015}. In this process, a protein efficiently searches for its ``target" site on the long DNA molecule and then binds the site strongly. To keep up with the metabolic, regulatory, reproductive, and defensive needs of the cell, these proteins must find their target sites sufficiently quick in some specific mechanism to avoid fruitless searching of target-less region. One strongly supporting evidence was observed by Riggs and co-workers \cite{Riggs1970} that \emph{in vitro} lac repressors find their target binding sites (operator) on \emph{E. coli} DNA at a surprisingly high rate measured as $10^{10}M^{-1}s^{-1}$, which demonstrates that the protein-DNA search is about 100 times faster than the Smoluchowski diffusion limit ($10^8 M^{-1}s^{-1}$) and 1000 times faster than protein-protein association ($10^6-10^7 M^{-1}s^{-1}$).
This inspired the idea of the ``facilitated diffusion" : the search process of protein consists of switching rounds of 3D diffusion in the bulk and 1D sliding diffusion along the DNA. The facilitated diffusion mechanism quantitatively explains the high protein-DNA association rates, as referred to in \cite{Berg1981},\cite{Berg1985},\cite{Slutsky2004},\cite{Kolomeisky2011}, and can be regarded as an optimal protein-DNA search mode in vivo.

Viewed as a dimensional reduction based on the linear topology of the DNA, two strategies are assumed to alternatively proceed to effectively accomplish the target searching process of the protein: the intersegmental jumps resulted from the association-dissociation event, and sliding diffusion along the DNA chain \cite{Coppey2004,Sheinman2012}. Physically, the intersegmental jumps effect a L\'evy flight (LF) search and the sliding diffusion corresponds to the Brownian motion. As an intuitive consideration, the LF with jumps can scan a large space avoiding the oversampling while the Brownian motion could exactly locate the target sites preventing the overshooting. It is the combination of these two motions that improves the searching efficiency significantly \cite{Koslover2011}. In fact, many biological processes in microscopic creatures have been proved to operate in a L\'evy style \cite{Kong1996, Shen2016}, such as the movement of \emph{E.coli} underlying their flagella rotation switching \cite{Korobkova2004, Tu2005} and the production of the regulatory proteins in the form of transcriptional bursts \cite{Xu2013, Zheng2016,Kang2017,Chen2017}. Generally, these processes are series of reaction-diffusion phenomena which can be expressed by the spacial fractional order Fokker-Plank equations. In terms of the target searching process, the protein slides on the DNA chain and combines to the target once the hydrogen-bond donor and acceptor groups of the target site on DNA match the peptide functional groups of  protein. The combination will lead to a probability loss of the protein occurring on the DNA which makes the total probability density non-normalized to unity. In physics, the searching problem comes down to the absorption of a particle undergoing the facilitated diffusion in the presence of a point sink, followed which the corresponding mathematical models can be extracted from either the gain-loss density or the diffusion reactions \cite{Halford2004,Mirny2009}. Based on these models, some analytical and numerical results are obtained \cite {Chechkin2003,Lomholt2005,Koren2007,Garcia2012,Janaliraman2017}. Chechkin et.al \cite{Chechkin2003} gave an estimation of the long-time behavior of the first arrival time density for free L\'evy flight with a perfectly absorbing sink. From the view of quantitativity, Lomholt et.al \cite{Lomholt2005}used the probability equation to describe the dissociation-association search process and obtained rich behaviors of the dynamics w.r.t. the physical parameters. Janakiraman \cite{Janaliraman2017} extended the results to arbitrary strength and position of the sink to obtain an analytical estimation of the absorption-time distribution.

However, as the sliding along the DNA chain is realized by the electrostatic interaction between the protein and the DNA backbone and the jumps between two segments finish in the nuclear matrix, the bias unavoidably occurs due to the change of the ion amounts in the solution or the interactions between the proteins and other macromolecules in nucleus. In particular, the presence of even a small bias which is on the direction towards the target location can lead to an optimal search strategy. Palyulin et.al \cite{JSM2014, Palyulin2014, Palyulin2017} introduced the constant drift velocity into the fractional Fokker-Planck equation for LFs to describe the bias occurring in LFs search process. Their observation demonstrated that the optimal search strategy varied with the bias and the initial position-target separation. In the case of the constant bias, they gave an analytical expression of the search efficiency as well as described the first arrival time density explicitly in the Laplace domain. Nevertheless, the bias occurs due to the external drift imposed on the protein, which is significantly influenced by the matrix environment and the position of the protein. So in general, facilitated diffusion model with the linear or nonlinear position-dependent bias is more practical other than with constant bias. In this case, as the inverse Laplace transform of the solution may not be performed analytically, an analytical solution in time domain cannot be obtained any more. It motivates us to employ some numerical methods to obtain the desired quantity for characterizing. Herein, we use a fast and accurate numerical algorithm developed  by Gao et al. \cite{Gao2016} to simulate the nonlocal Fokker-Planck equation with natural boundary condition. This applies to stochastic systems with finite as well as small noise intensity.

\indent The paper is organized as follows. In section 2, we present a brief introduction on the biological background of the target search model and the corresponding nonlocal Fokker-Planck equation to describe the density evolution of the proteins. In section 3, we qualify the searching mechanisms, by a numerically solving the Fokker-Planck equation and then calculating two important indexes, survival probability and first arrival density. In this section, both linear and nonlinear biases are taken into consideration and the roles of the parameters in the model are examined. Finally, we summarize the results and conclude with a brief discussion in section 4.

\section{Model}
\begin{figure}
\centering
\includegraphics[height = 7cm, width = 10cm]{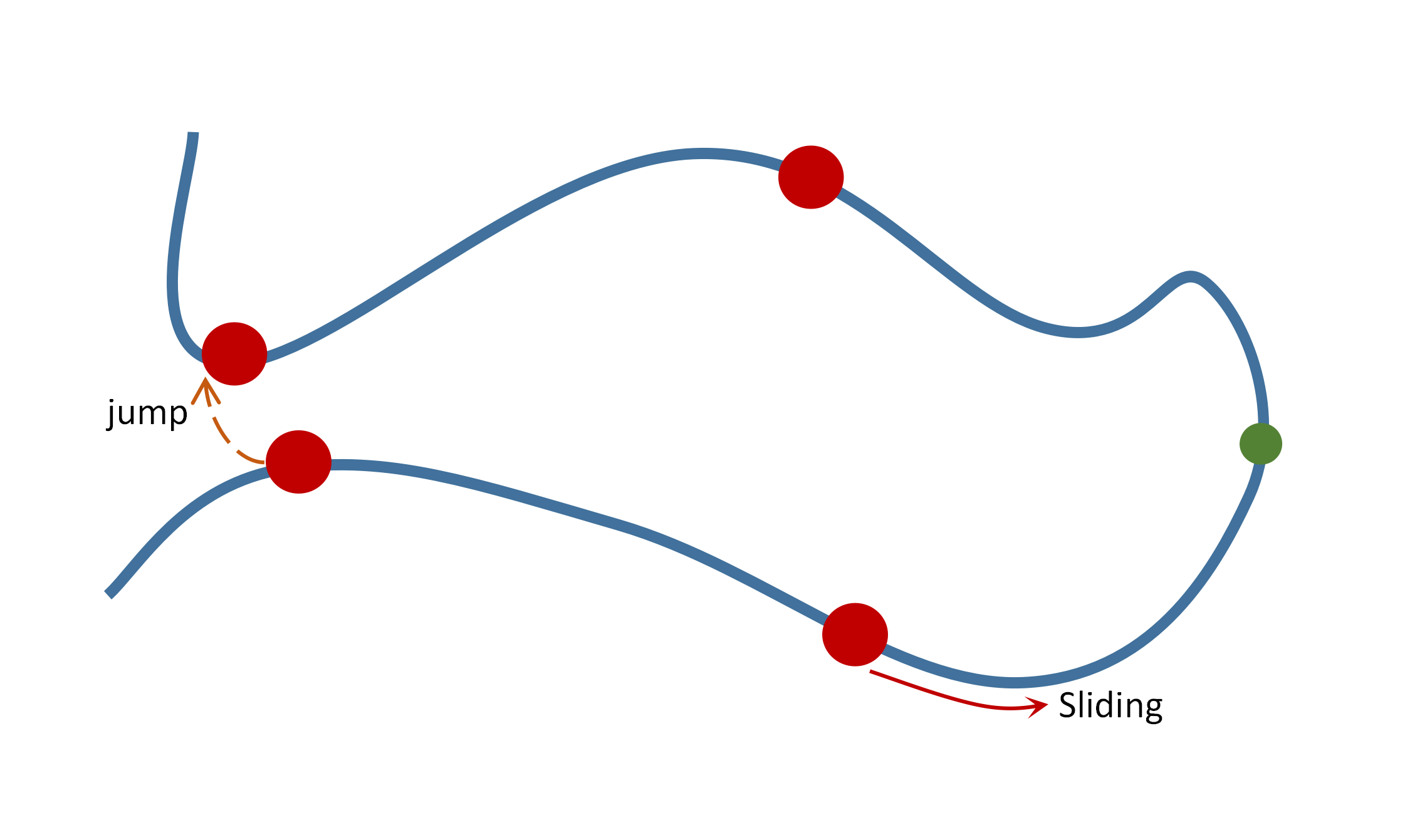}
\caption{Target search of a protein along a fast-folding DNA chain by facilitated diffusion. The red ball represents the protein molecular and the green ball denotes the target site on the DNA. The blue line is the DNA chain.}
\label{Fig1}
\end{figure}

In the description of the target search process, we use the quantity $p(x,t)$ to represent the possibility of finding the protein at position $x$ at time $t$, which can be considered as a possibility density function (PDF) \cite{Lomholt2005}. Assuming that at $t=0$ the protein is located at $x=x_0$, we have the initial condition $p(x,0)=\delta(x-x_0)$. As referred in the previous section, the searching process includes sliding along the DNA and intersegmental jumping among the DNA segments, which can be mathematically expressed by the Brownian motion and L\'evy motion respectively. Apart from these two diffusions, we take the bias (drift velocity) into consideration. Thus the dynamics of $p(x,t)$ is derived from the Fokker-Planck equation \cite{Metzler2000,Metzler2004}
\begin{eqnarray}
\frac{\partial}{\partial t}p(x,t)=d\frac{\partial ^{2} p(x,t)}{\partial x^{2}}+\epsilon\frac{\partial ^{\alpha}p(x,t)}{\partial |x|^{\alpha}}-\frac{\partial (v(x)p(x,t))}{\partial x}-\phi(t)\delta(x-x_s).
\label{equation1}
\end{eqnarray}
In this equation, $d$ is the diffusion constant related to the sliding motion on the DNA and $\epsilon$ is the non-negative noise intensity characterizing intersegmental jumps. $v(x)$ denotes the drift, and the term $\phi(t)\delta(x-x_s)$ represents the target sinking at $x=x_s$. The function $\phi(t)$ describes the flux into the target: in terms of the whole system, once the random walker arrives at the target, it is removed immediately. That is to say, the target is perfectly absorbing and $\phi(t)$ is physically defined as the density of first arrival. Integrating Eq.(\ref{equation1}) over the state space delivers the expression of the first arrival density:
\begin{equation}
\phi(t)=  -\frac{d}{dt}\int_{-\infty}^{\infty}p(x,t)dx.
\end{equation}
Owing to the sink term, $p(x,t)$ is not normalized. So the cumulative survival probability (not hitting the target up to time t)
\begin{equation}
P(t)=\int_{-\infty}^{\infty}p(x,t)dx
\end{equation}
is a deceasing function of time and $\phi(t)$ is always non-negative. Different from the consideration of constant bias in \cite{Palyulin2014}, the drift term $v(x)$ in Eq.(\ref{equation1}) can be a position-dependent function varied by the practical matrix density. Moreover, the fractional derivative $\frac{\partial^{\alpha}}{\partial |x|^{\alpha}}$ in Eq.(\ref{equation1}) is equivalent to the fractional Laplacian operator $-(-\triangle)^{\frac{\alpha}{2}}$, which can be defined as \cite{Applebaum2009, Duan2015,Schertzer2001}
\begin{equation}
-(-\triangle)^{\frac{\alpha}{2}}p(x,t)=\int_{\mathbb{R}\backslash{\{0\}}}[p(x+y,t)-p(x,t)-I_{\{|y|<1\}}yp_{x}(x,t)]\nu_{\alpha}(dy).
\label{equation2}
\end{equation}
This is the generator of the $\alpha$-stable L\'evy motion and $\nu_{\alpha}(dy)$ is the jump measure with definition

$$\nu_{\alpha}(dy)=C_{\alpha}\left|y\right |^{-(1+{\alpha})}dy, \mbox{ where} \ C_{\alpha}=\frac{\alpha}{2^{1-\alpha}\sqrt \pi}\frac{\Gamma (\frac{1+\alpha}{2})}{\Gamma(1-\frac{\alpha}{2})}.$$
Here $\Gamma$ is the Gamma function. $\alpha\in(0,2)$ is called the non-Gaussianity index or stable index, which characterizes the non-Gaussianity of the  $\alpha$-stable L\'evy motion. Intuitively speaking, the smaller the $\alpha$ index is, the more prominent the non-Gaussianity or heavy-tailed property is. Specifically to the target search process, it has been proved that L\'evy search for a point-like target will never succeed for $0<\alpha\leq1$, so in this paper we consider the non-Gaussianity in the range $1<\alpha<2$. When $\alpha=2$, the symmetric $\alpha$-stable process is simply the Brownian motion. We can further rewrite Eq.(\ref{equation1}) as

\begin{eqnarray}
\partial_{t}p(x,t)=&-\partial_{x}(v(x)p(x,t))+d\partial_{xx}p(x,t)-\phi(t)\delta(x-x_s)\nonumber\\
&+\varepsilon\int_{\mathbb{R}\backslash{\{0\}}}[p(x+y,t)-p(x,t)-I_{\{|y|<1\}}yp_{x}(x,t)]\nu_{\alpha}(dy).
\label{equation3}
\end{eqnarray}

In the target searching process, the proteins may locate arbitrary sites on the DNA. So here we consider the problem in the case of natural boundary condition: the DNA chain is so long that one can set the boundary of stretched DNA chain at infinity, which means the PDF at infinity is zero. The technique of extrapolation is adopted here to meet the natural boundary condition in our algorithm. To reduce the calculating cost, we lay a truncation on the original infinite domain based on an convergence discussion \cite{Gao2016}.
As a theoretical support, the solution of the non-local Fokker-Planck equation has been proved to exist in $L^{p}$ for $p\geq1$ when the drift is linear and double-well \cite{Wang2017}. So in our investigation, we numerically calculate the solution of Eq.(\ref{equation3}) in the case of linear and double-well drift terms, which cover interesting situations that evoke biases.

We use a similar efficient numerical finite difference method developed by Gao et al. \cite{Gao2016} to simulate the non-local Fokker-Planck  equation Eq.(\ref{equation3}). As we get the PDF of the protein located at $x$ on time $t$, we can further introduce the search reliability $R(t)$ up to time $t$ to characterize the search mechanism \cite{Palyulin2014,Palyulin2017}:
\begin{equation}
R(t)=\int_{0}^{t}\phi(\xi)d\xi=1-P(t).
\end{equation}
Owing to the definition of $R(t)$ as the integral of first arrival density, it quantifies the probability that the protein ever arrives at the target up to time $t$. So a large value of $R(t)$ corresponds to a high success probability of locating the target, which has the same indication of a small survival probability $P(t)$. Substantially, they are the quantities explaining the same thing from two sides and one can choose one of them to describe the search efficiency alternatively. In our numerical experiments below, we use the survival probability $P(t)$ to characterize the searching efficiency. As for the first arrival density which depicts the variation rate of the search reliability, we can use it to quantify the search efficiency. Based on the survival probability and first arrival density, the properties of the search strategies can be analyzed from the view of quantitativity.

\section{Results}

\subsection{Target search with linear drift: $v(x)=kx$}

\indent As is known, the motions of the proteins finding their specific targets are usually taken place in the nuclear matrix and the density of the matrix is closely related to the division period of the cells. In this part we assume the environment of the nuclear matrix is not crowded, which means the block caused by the macromolecules on the moving process could be ignored and the only resistance comes from the viscous drag of the solution. So we can treat the external force as the linear drift. Starting with the case of linear drift $v(x)=-0.01x$, the PDF of the proteins locating on the DNA is numerically obtained from the Fokker-Planck equation Eq.(\ref{equation3}) as well as other important measures to quantify the search mechanism. Since the search mechanism is substantially related to the non-Gaussianity and the initial location of the protein \cite{Palyulin2014,Palyulin2017}, we compare the search reliability and efficiency (i.e.,  the  first arrival density) for different non-Gaussianity indexes and initial locations $x_0$.

\indent The PDF of the searching proteins with linear drift and different $\alpha$-stable index is shown in Figures 2-4. It can be seen that the possibility density of the proteins firstly concentrates on the initial location, which means the proteins start to move but are restricted to the region around the initial point. As time goes on, with the help of the electrostatic and external drift force, the proteins locate more and more DNA segments both in the motion of sliding and jumps, so the possibility of the proteins occurring at different locations increases. Interestingly, the PDF curve will ``drop" at the target site $x=1$ after some time instant, which infers that the proteins will be absorbed or removed at this location. Correspondingly, the total probability flux will flow away through the target as time evolves. The decreasing of the survival probability with time in Figures 2(b)-4(b) further supports this behavior. The results appear to be consistent with a convection-diffusion equation with $\delta$-sink: when a sink is introduced into a dynamical system, the flux will flow into the sink which leads to an ``inflection'' at the sink.

\begin{figure}[H]
\centering
\subfigure[]{
\centering
\includegraphics[height = 6cm, width = 10cm]{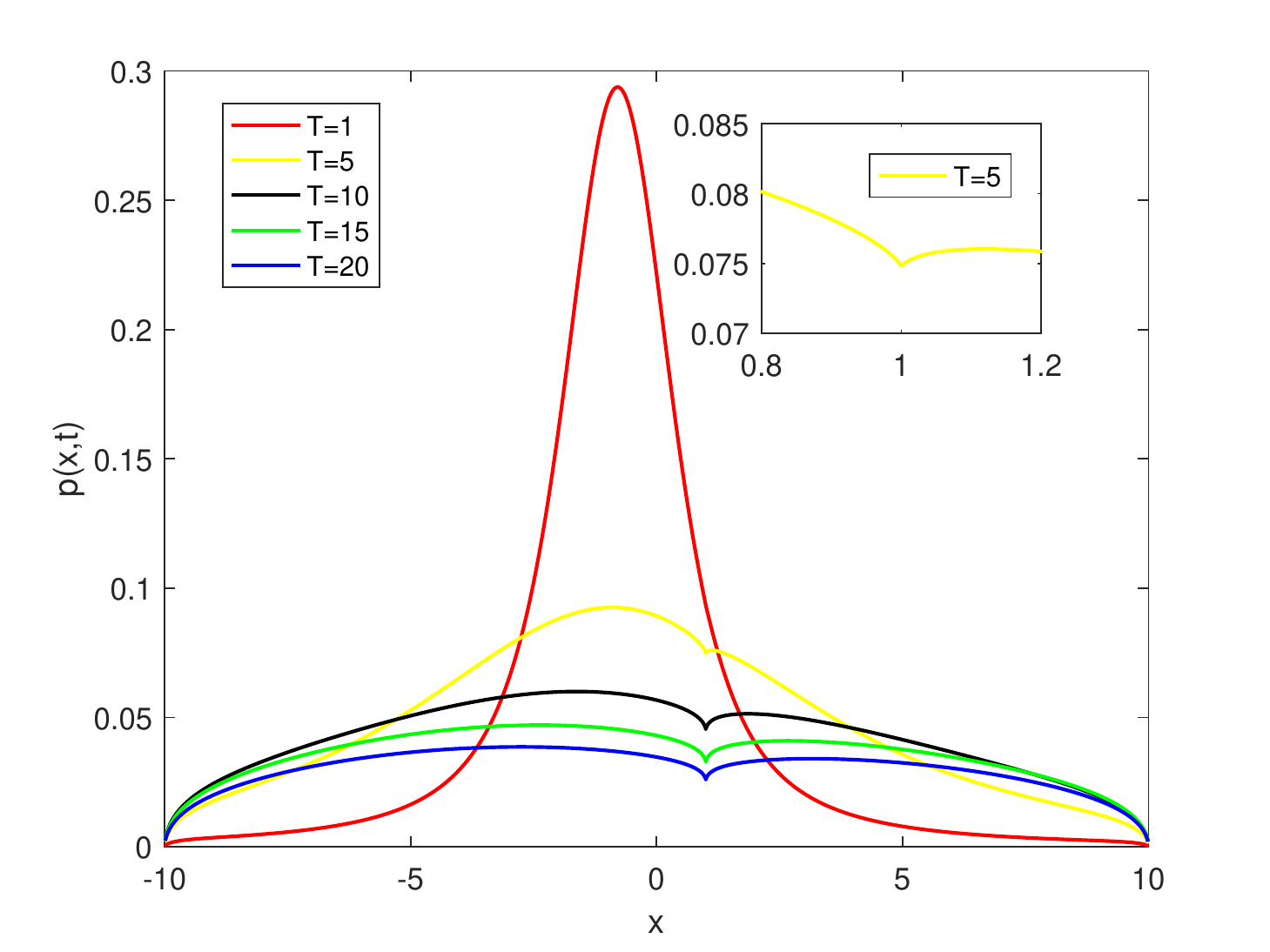}}
\subfigure[]{
\raggedright
\includegraphics[height = 6cm, width = 6cm]{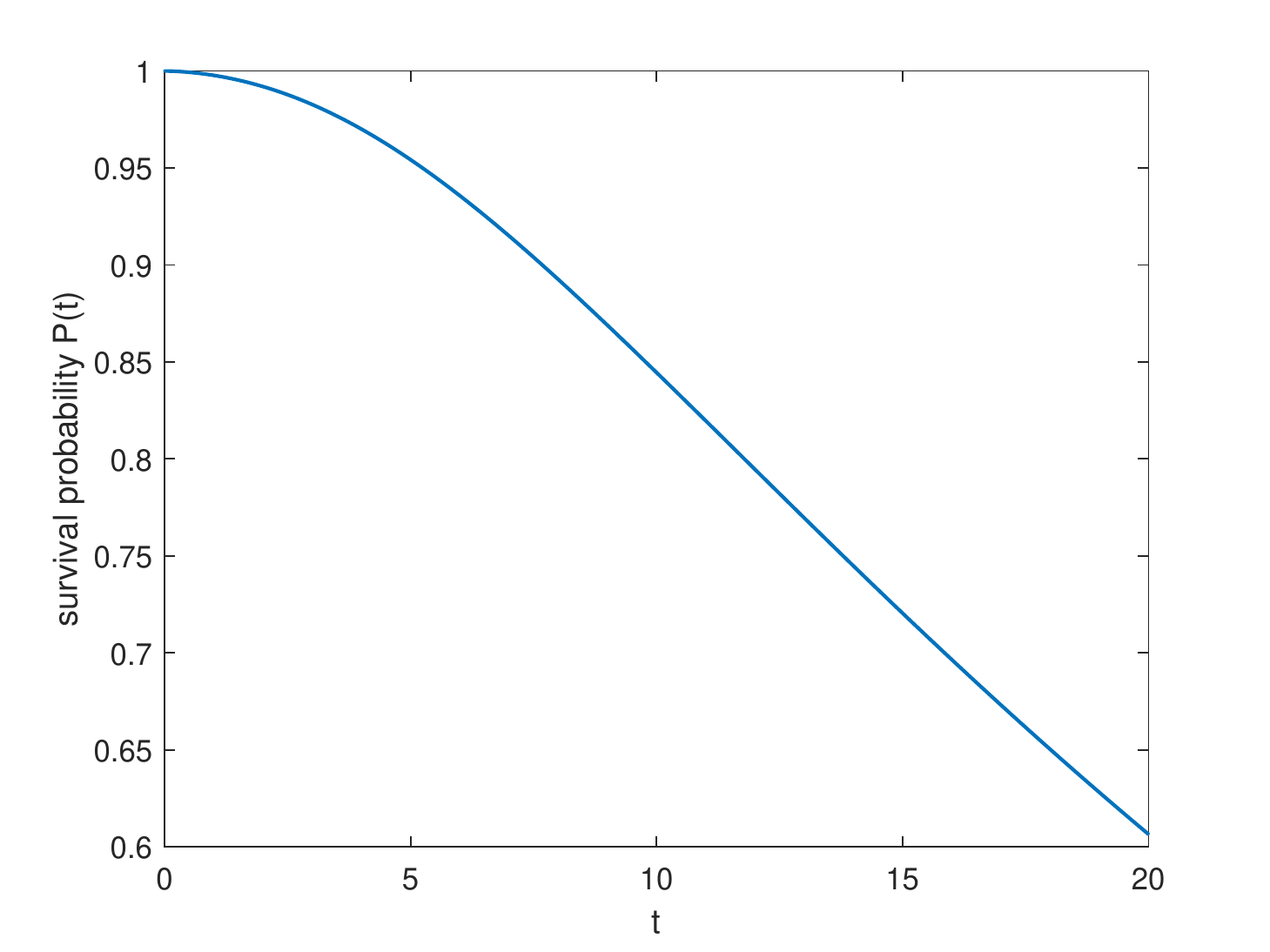}}
\subfigure[]{
\raggedleft
\includegraphics[height = 6cm, width = 6cm]{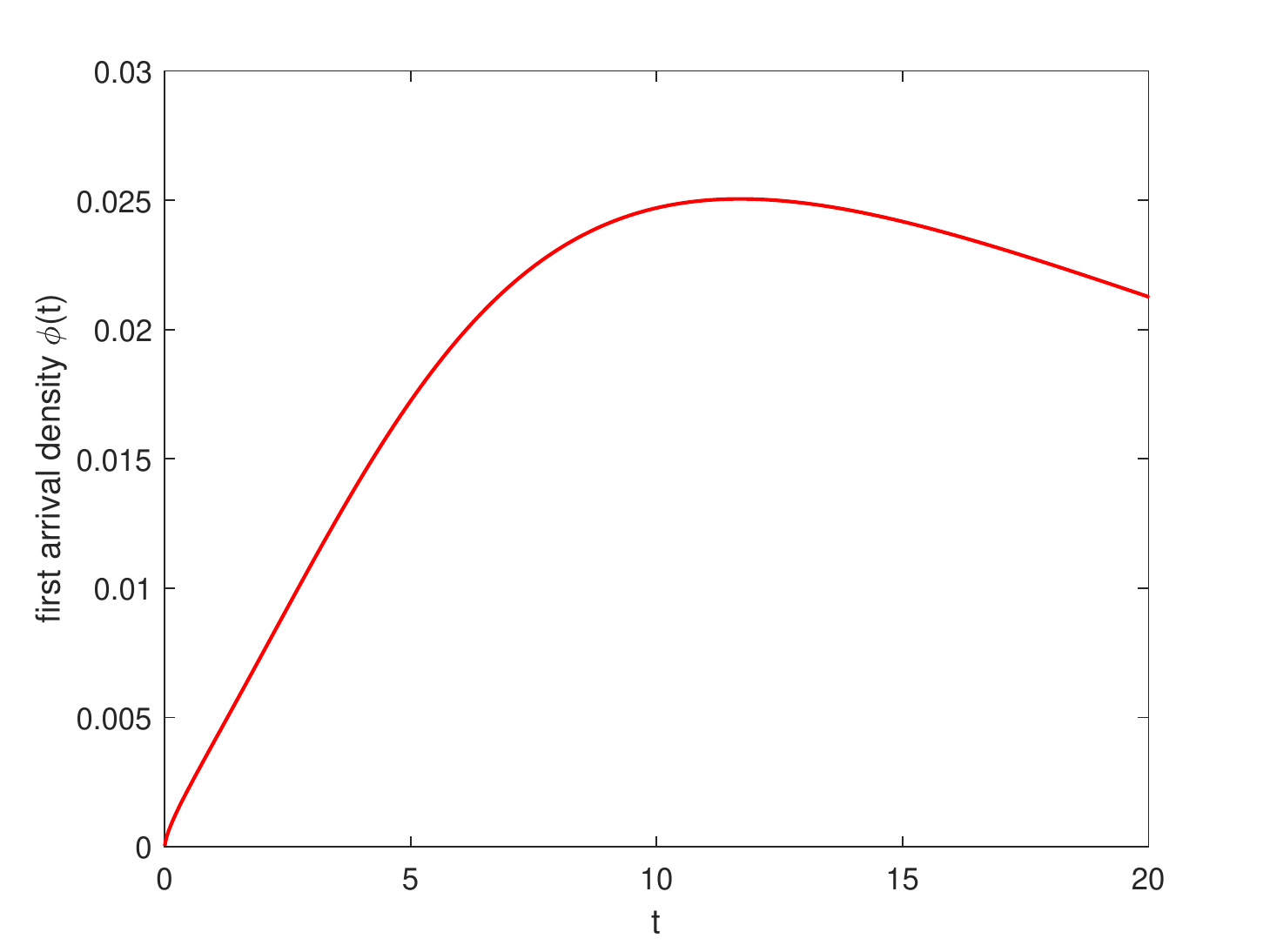}}
\centering
\caption{The search strategy in the linear drift case $v(x)=-0.01x$ with non-Gaussian index $\alpha=1.2$,  the initial location at $x_0=-0.8$,  the target location at $x_s=1$,  noise intensity $\epsilon=1$ and diffusion constant $d=0.1$:
(a) The curves of PDF at different time instants.  (b) The survival probability up to time $t$.  (c) The first arrival density $\phi(t)$. }
\label{Fig2}
\end{figure}

\begin{figure}[H]
\centering
\subfigure[]{
\centering
\includegraphics[height = 6cm, width = 10cm]{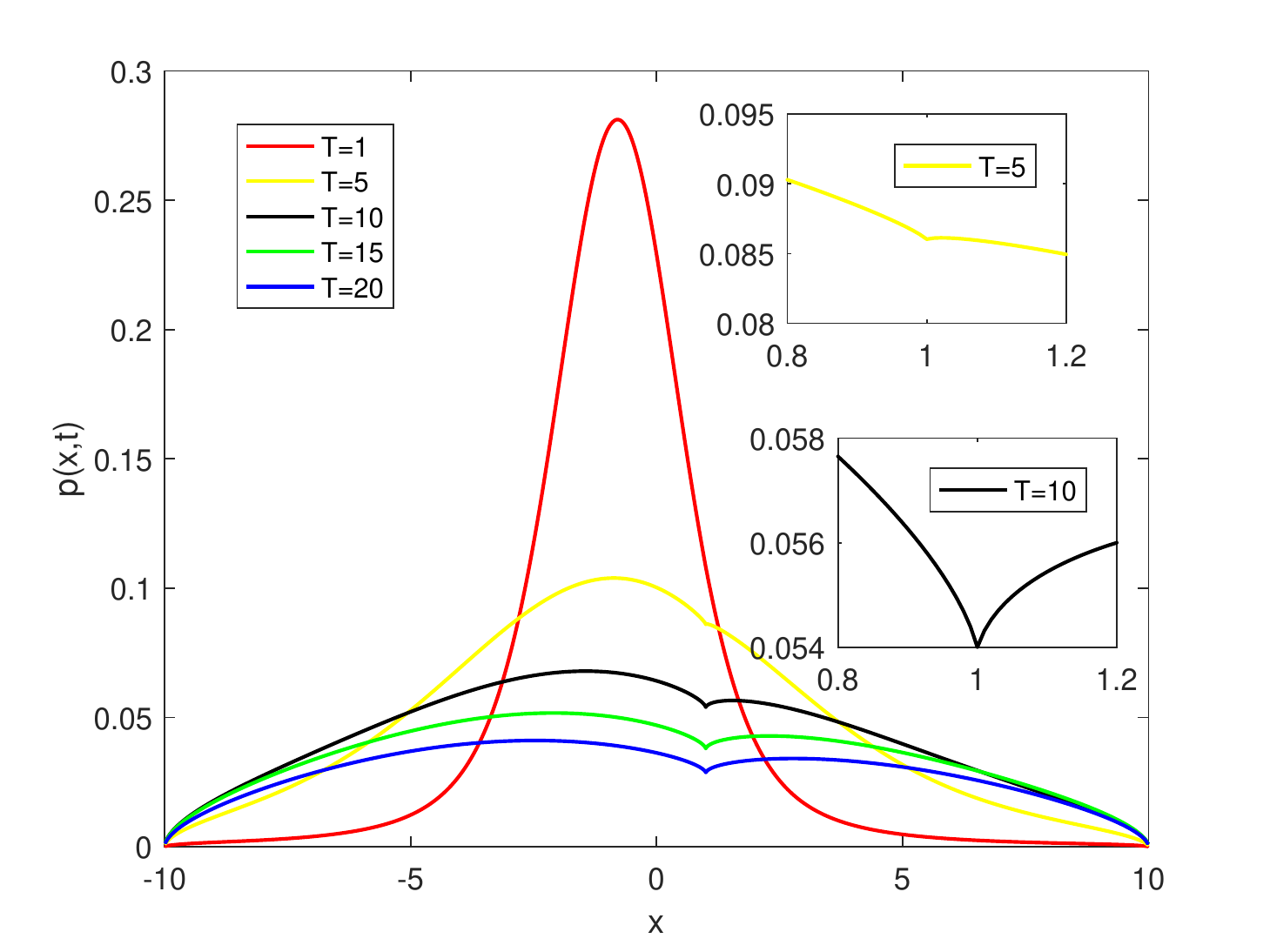}}
\subfigure[]{
\raggedright
\includegraphics[height = 6cm, width = 6cm]{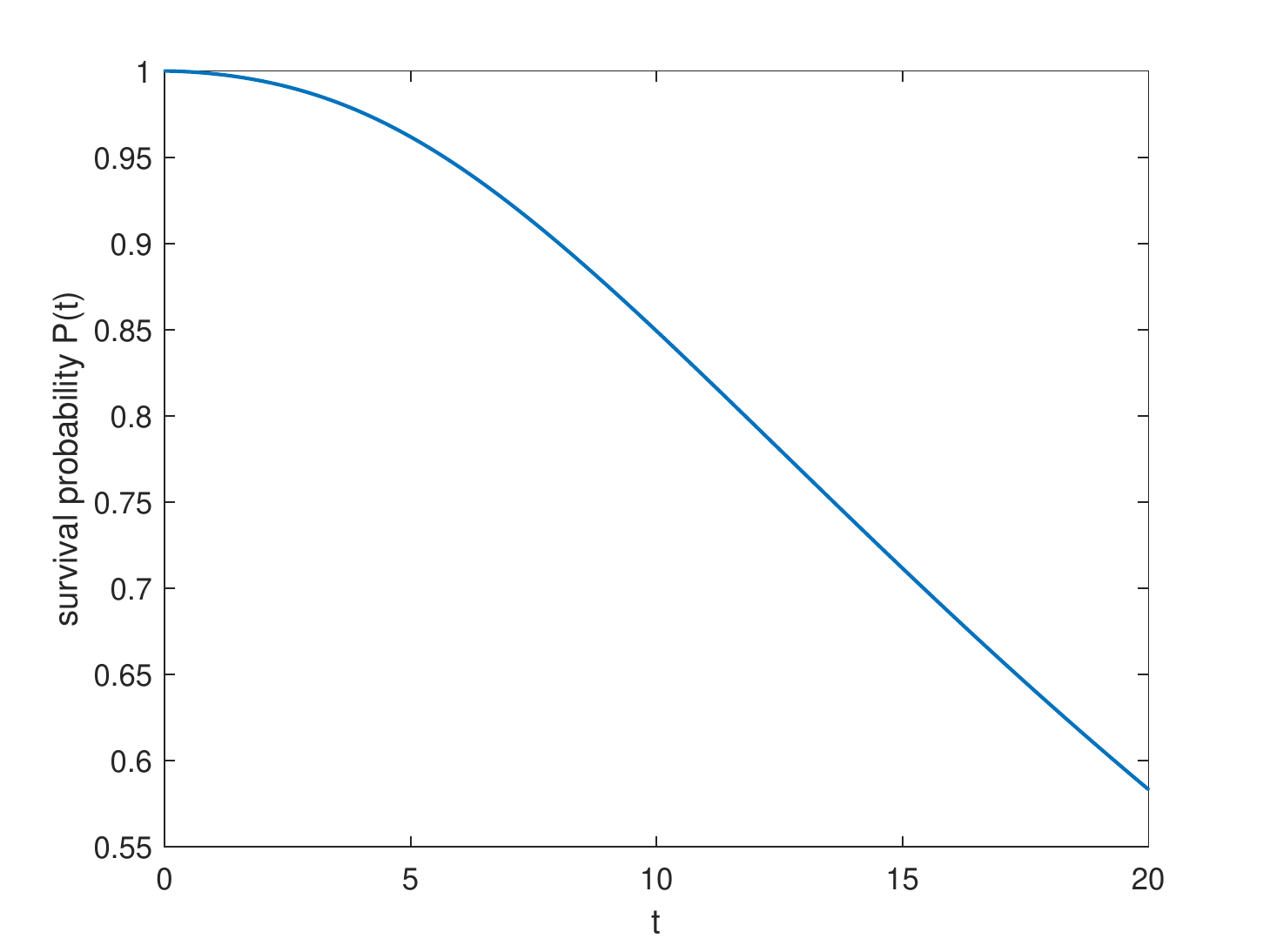}}
\subfigure[]{
\raggedleft
\includegraphics[height = 6cm, width = 6cm]{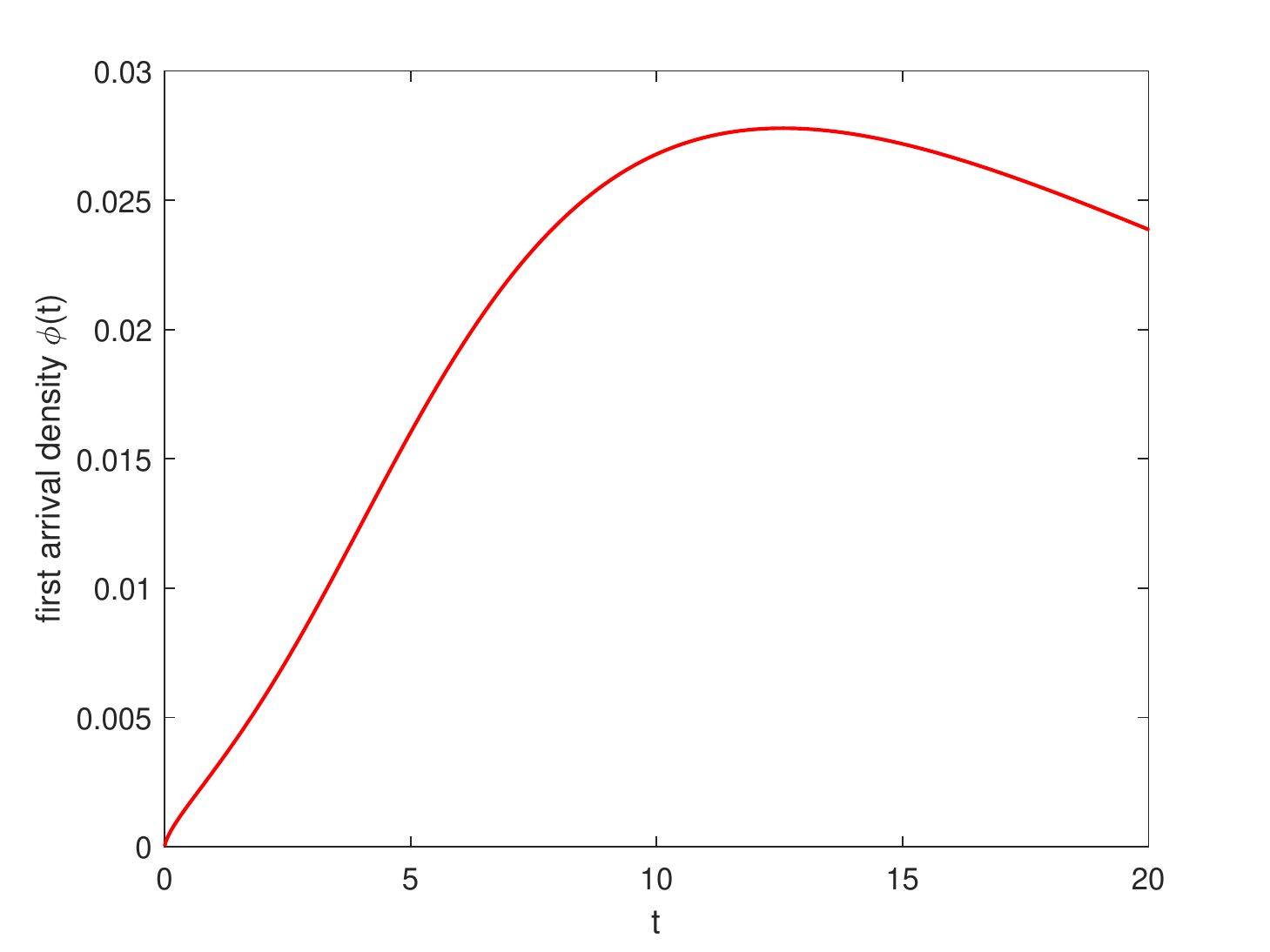}}
\caption{The search strategy in the linear drift case $v(x)=-0.01x$ with non-Gaussian index $\alpha=1.5$,  the initial location at $x_0=-0.8$,  the target location at $x_s=1$,  noise intensity $\epsilon=1$ and diffusion constant $d=0.1$:
(a) The curves of PDF at different time instants.  (b) The survival probability up to time $t$.  (c) The first arrival density $\phi(t)$. }
\label{Fig3}
\end{figure}

\begin{figure}[H]
\centering
\subfigure[]{
\centering
\includegraphics[height = 6cm, width = 10cm]{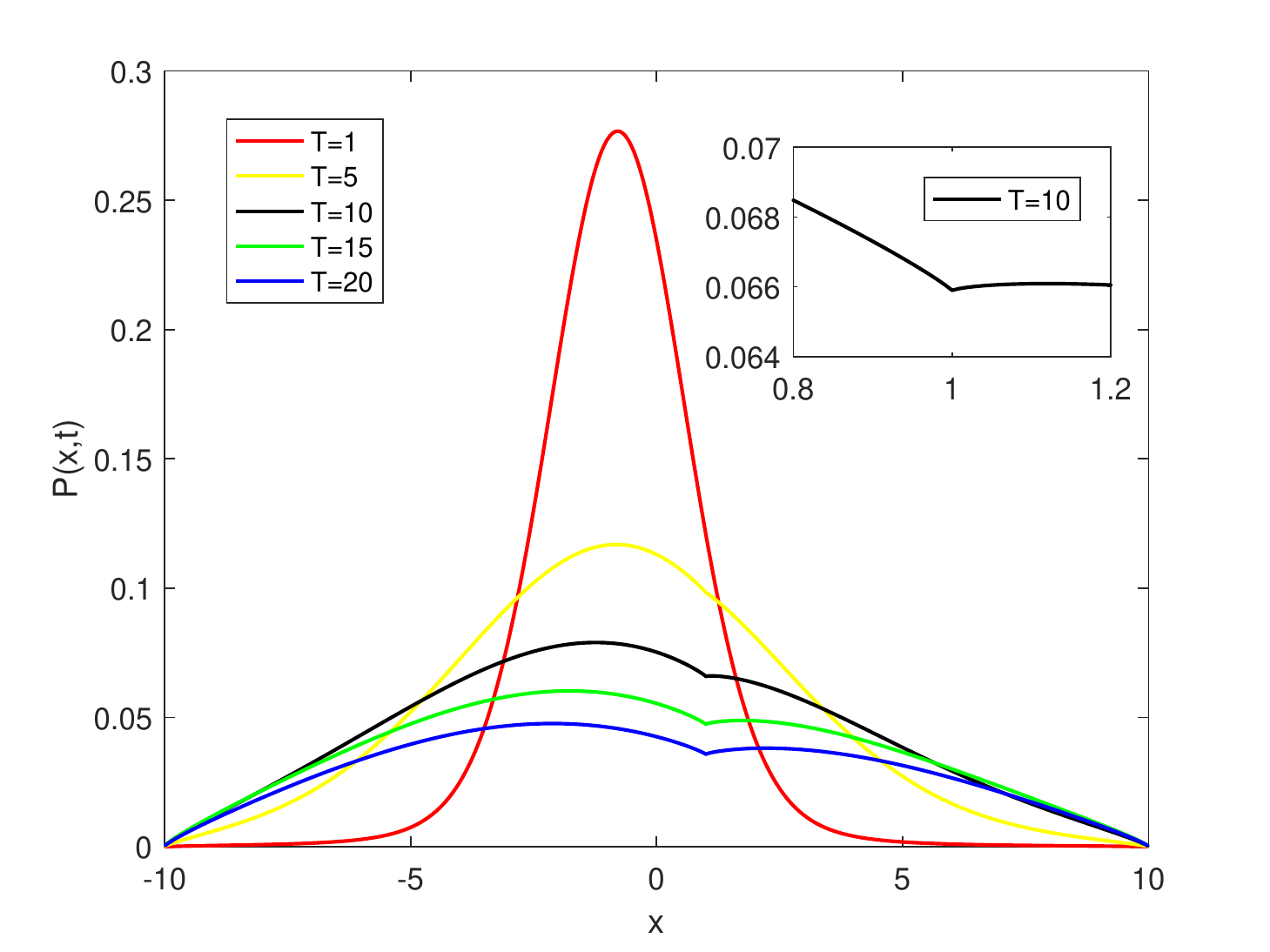}}
\subfigure[]{
\raggedright
\includegraphics[height = 6cm, width = 6cm]{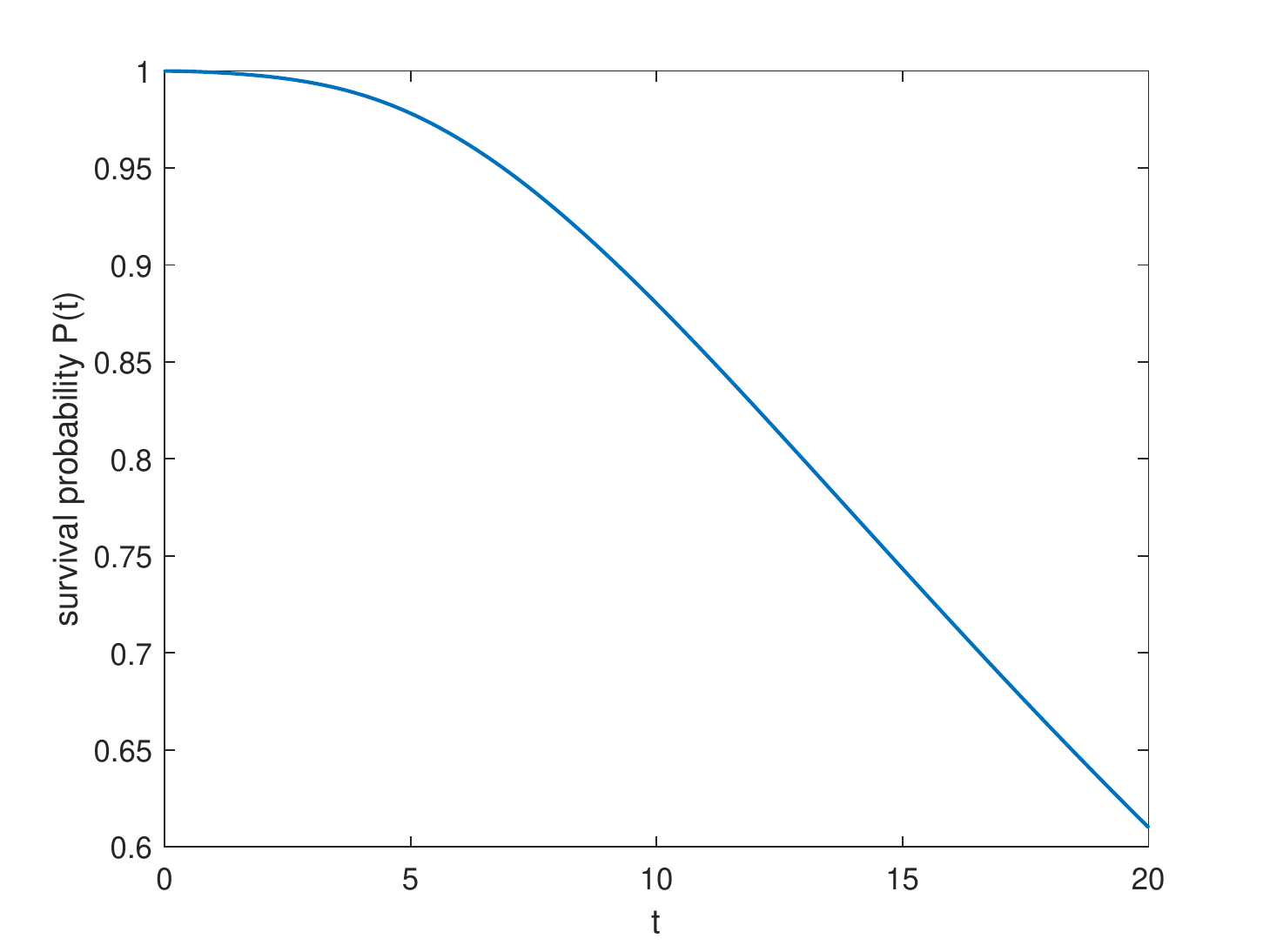}}
\subfigure[]{
\raggedleft
\includegraphics[height =6cm, width = 6cm]{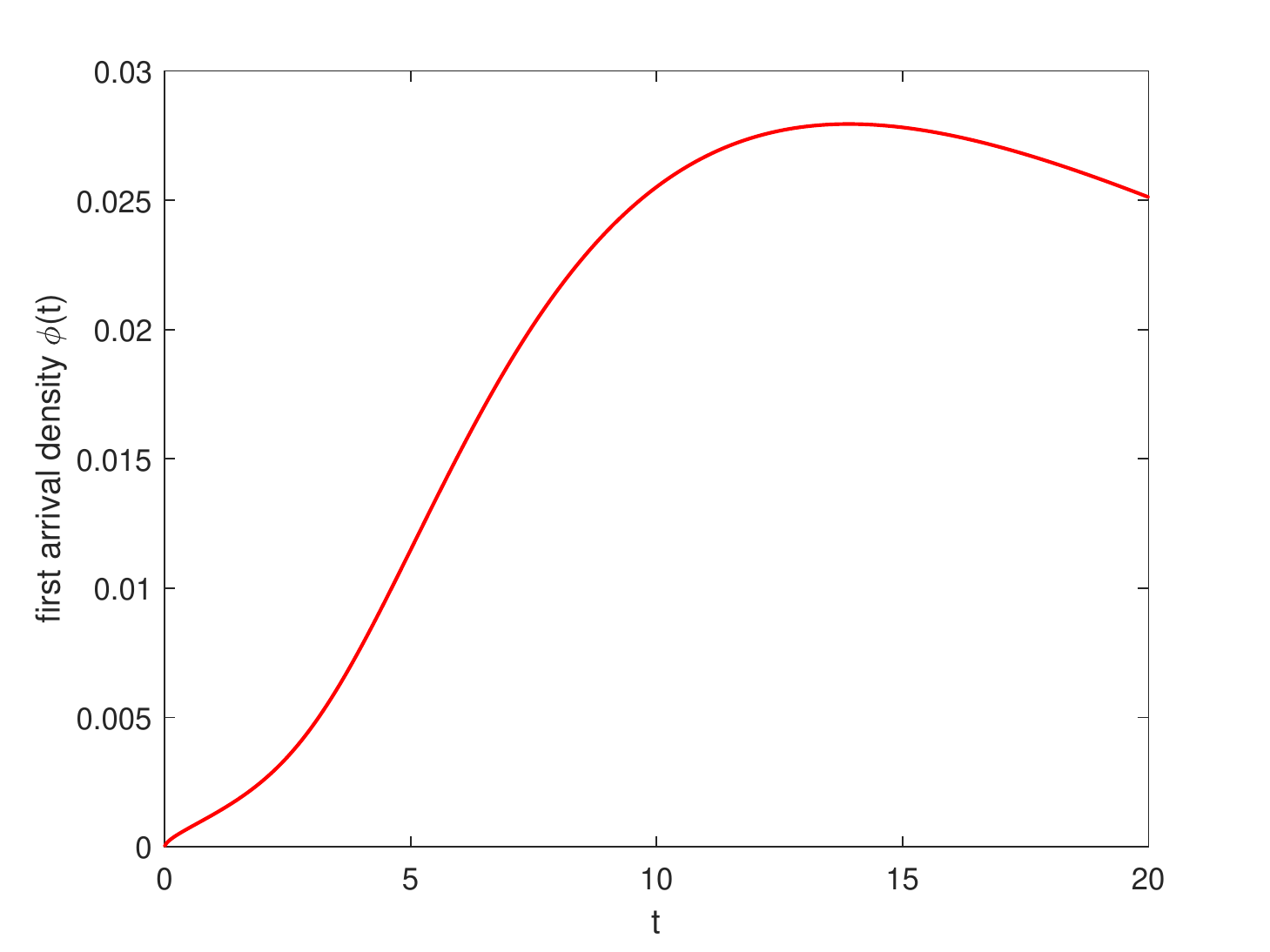}}
\caption{The search strategy in the linear drift case $v(x)=-0.01x$ with non-Gaussian index $\alpha=1.8$,  the initial location at $x_0=-0.8$,  the target location at $x_s=1$,  noise intensity $\epsilon=1$ and diffusion constant $d=0.1$:
(a) The curves of PDF at different time instants.  (b) The survival probability up to time $t$.  (c) The first arrival density $\phi(t)$.}
\label{Fig4}
\end{figure}

\begin{figure}[H]
\centering
\subfigure[]{
\includegraphics[height = 7.5cm, width = 6cm]{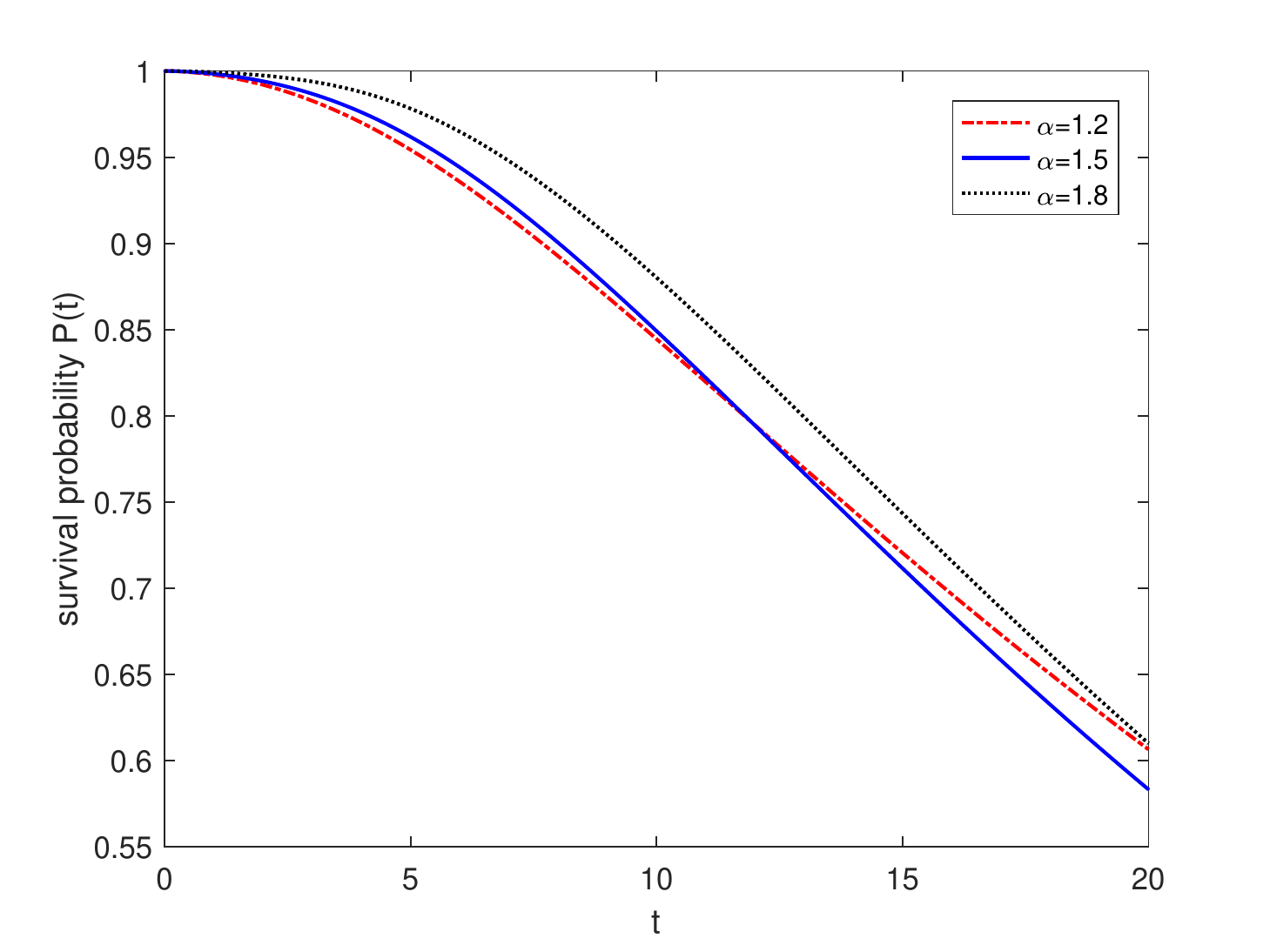}}
\subfigure[]{
\includegraphics[height = 7.5cm, width = 6cm]{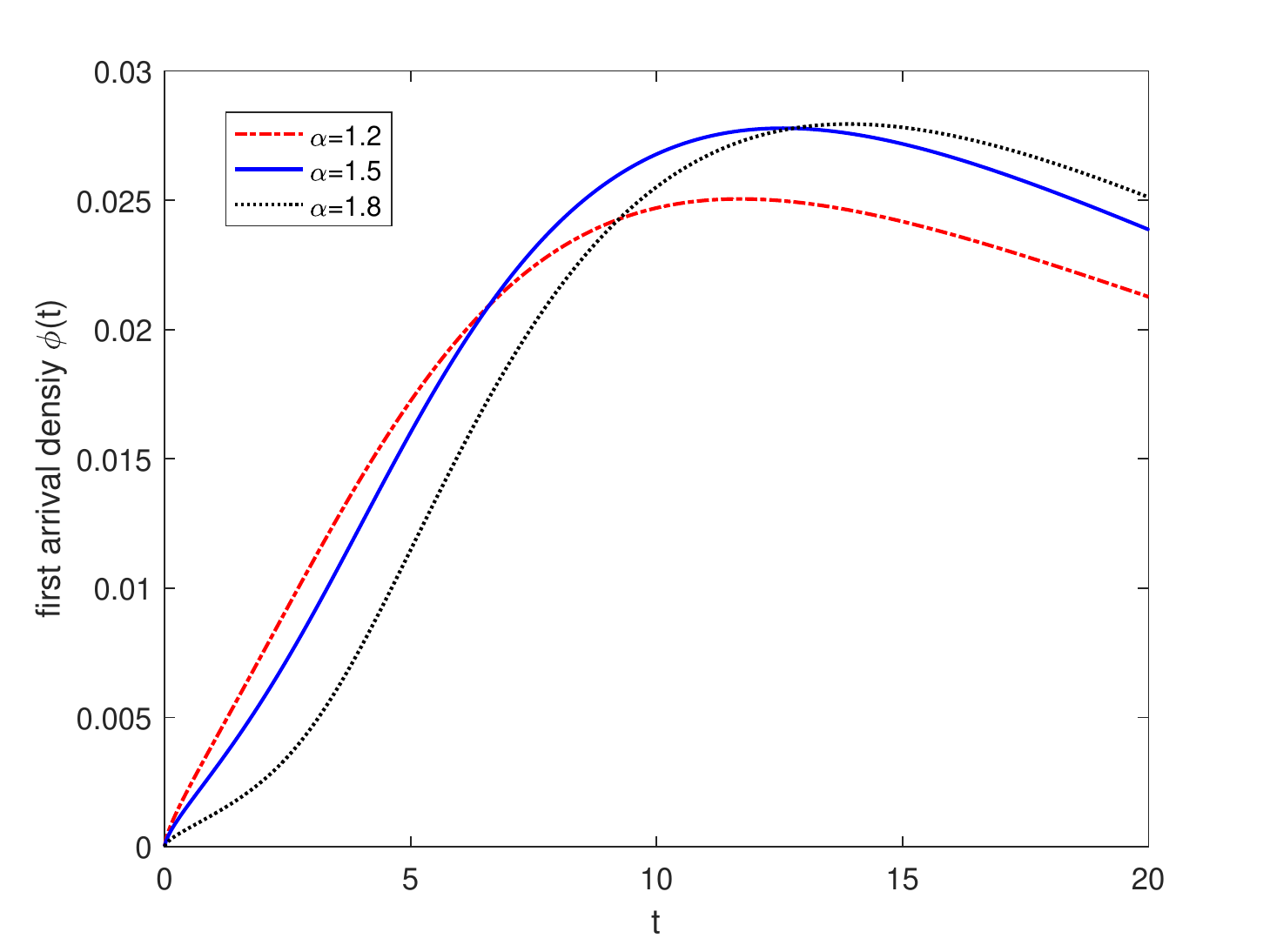}}
\caption{The comparisons of the survival probability and first arrival density with different non-Gaussianity indexes $\alpha$ for the linear drift case $v(x)=-0.01x$. Here the initial location is $x_0=-0.8$ and the target locates on $x_s=1$. The noise intensity is $\epsilon=1$. }
\label{Fig5}
\end{figure}

\indent Moreover, we observe from Figure 5 that when the initial point is fixed at $x=-0.8$, the``inflection''appears earlier with a smaller non-Gaussianity index $\alpha$ than a larger $\alpha$ index case. However, by comparing the survival probability and the first arrival density with different $\alpha$ indexes, one can see that the relationship between the non-Gaussian $\alpha$ index and the searching reliability or efficiency is not monotonically-dependent. The search strategy with more non-Gaussionity at first leads to a higher search reliability, but as time goes on, its superiority is lost to the other search mechanism with a larger $\alpha$ index. So we can suggest that after long enough time evolution or in the stationary state, for the fixed initial location, there should be an optimal $\alpha$ belonging to $(1,2)$ that makes the survival probability minimum, i.e. a maximum search reliability.

\begin{figure}[H]
\centering
\subfigure[]{
\centering
\includegraphics[height = 6cm, width = 10cm]{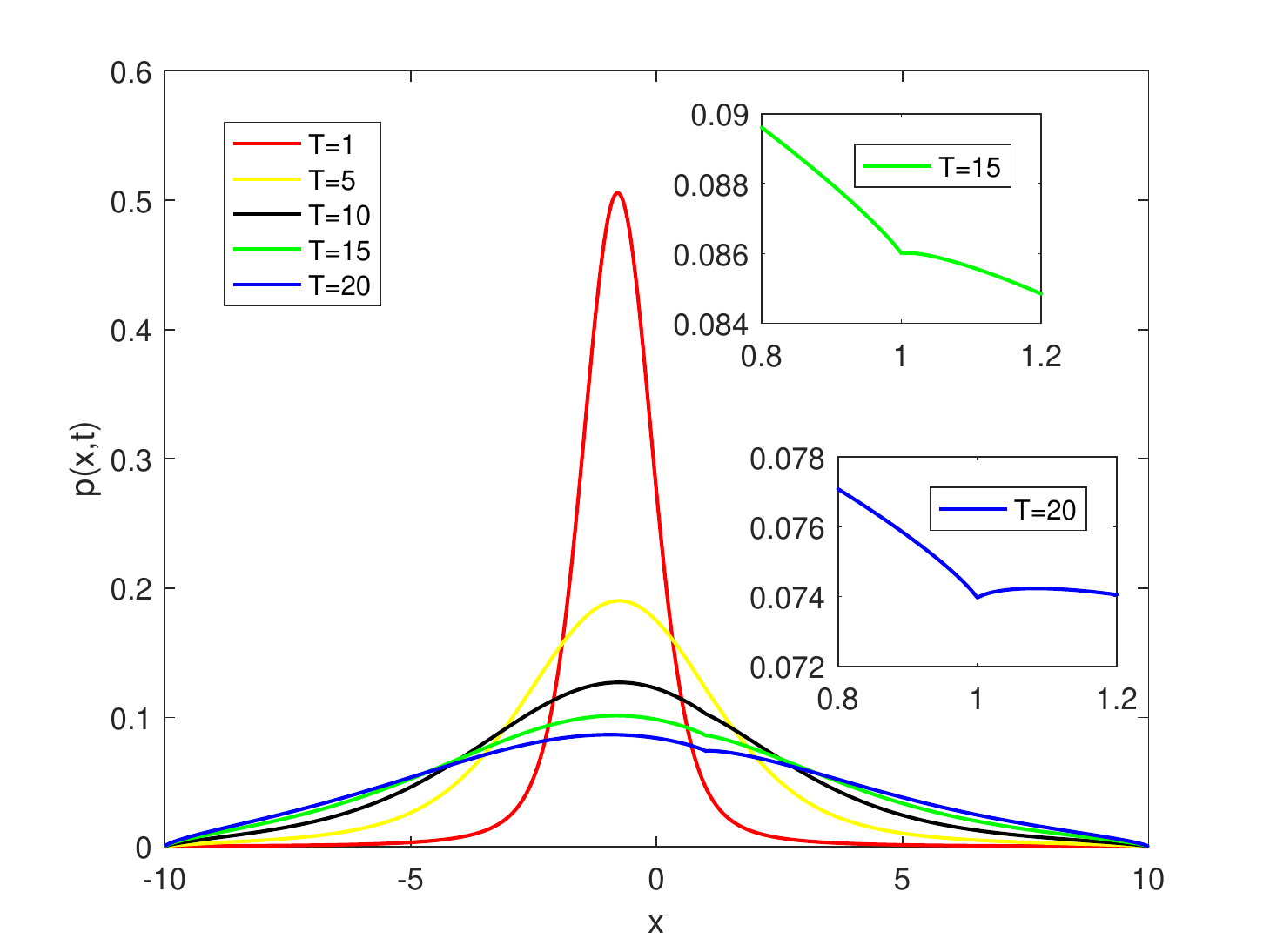}}
\subfigure[]{
\raggedright
\includegraphics[height = 6cm, width = 6cm]{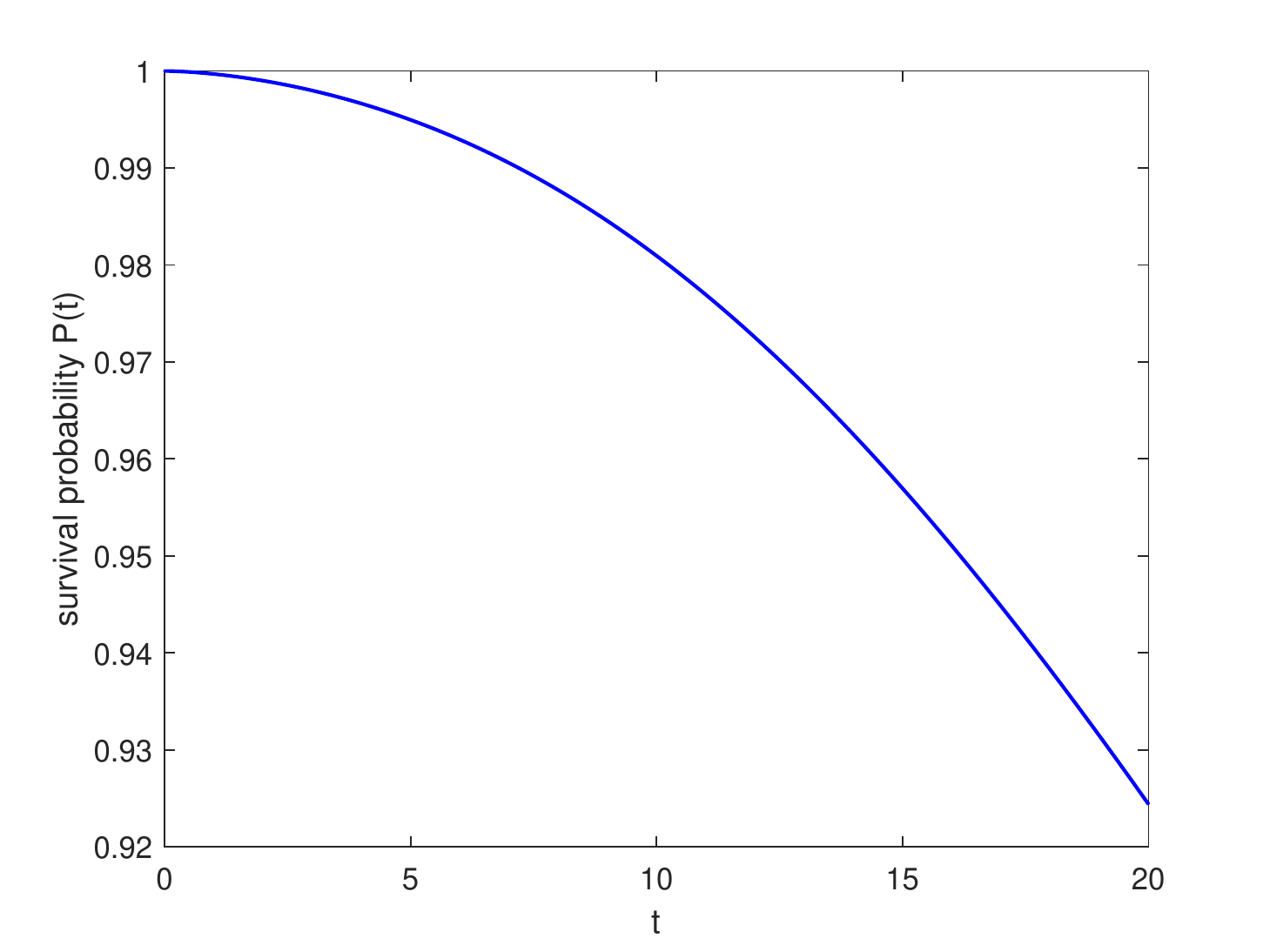}}
\subfigure[]{
\raggedleft
\includegraphics[height = 6cm, width = 6cm]{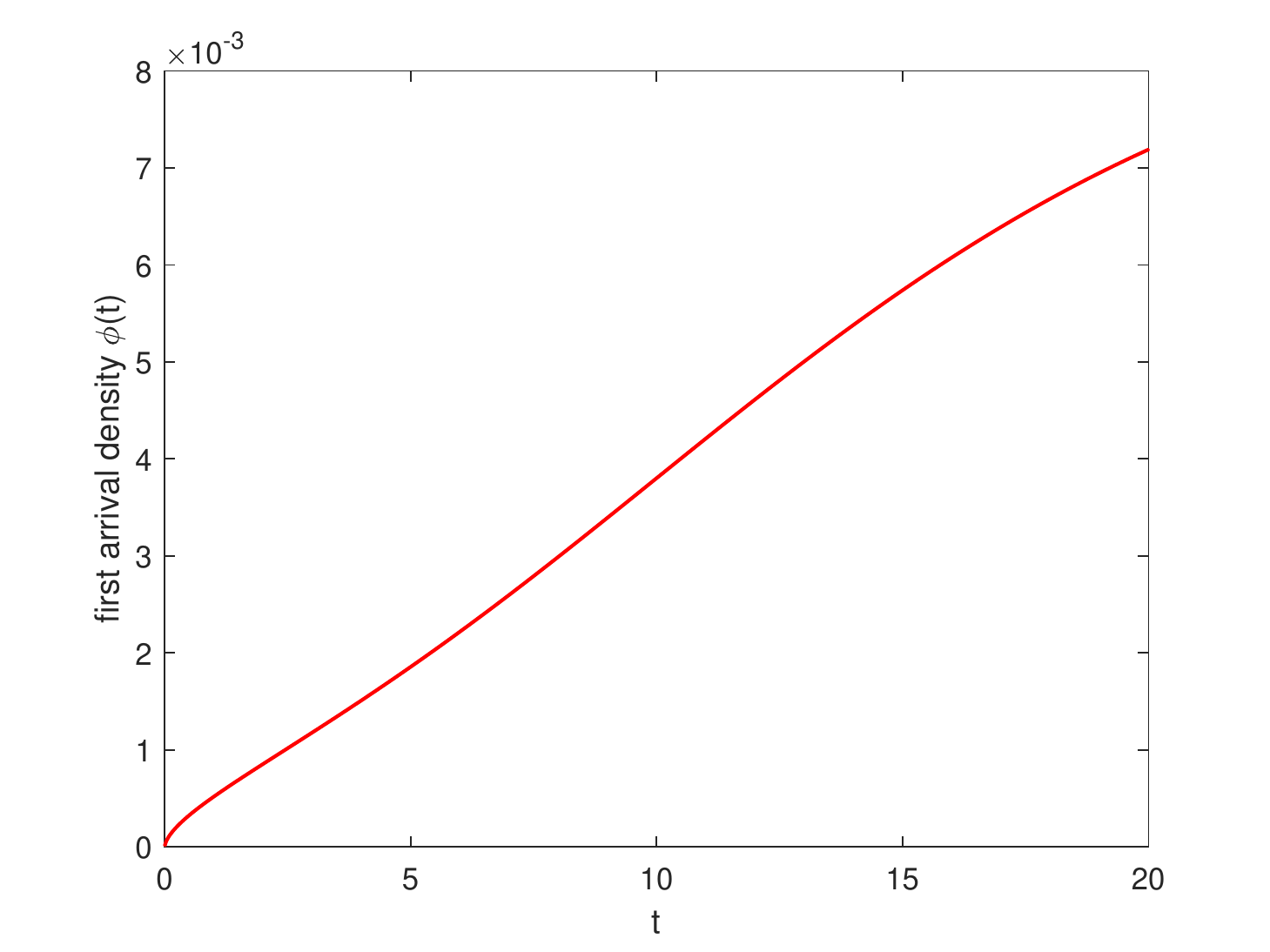}}
\caption{The search strategy in the linear drift case $v(x)=-0.01x$ with noise intensity $\epsilon=0.5$, the initial location at $x_0=-0.8$, the target location at $x_s=1$, non-Gaussianity index $\alpha=1.5$ and diffusion constant $d=0.1$:
(a) The curves of PDF at different time instants.  (b) The survival probability up to time $t$.  (c) The first arrival density $\phi(t)$. }
\label{Fig6}
\end{figure}

\begin{figure}[H]
\centering
\subfigure[]{
\includegraphics[height = 7.5cm, width = 6cm]{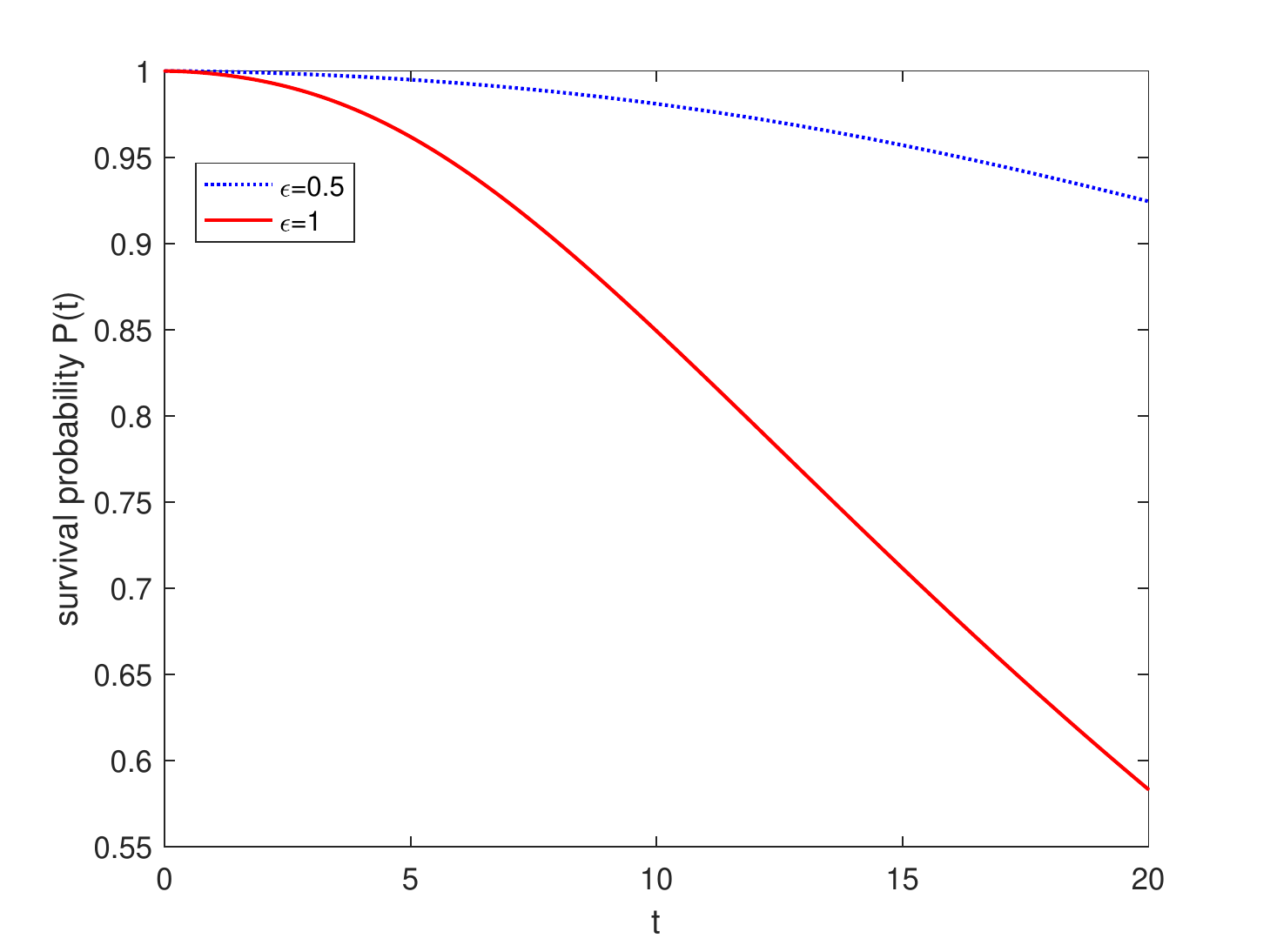}}
\subfigure[]{
\includegraphics[height = 7.5cm, width = 6cm]{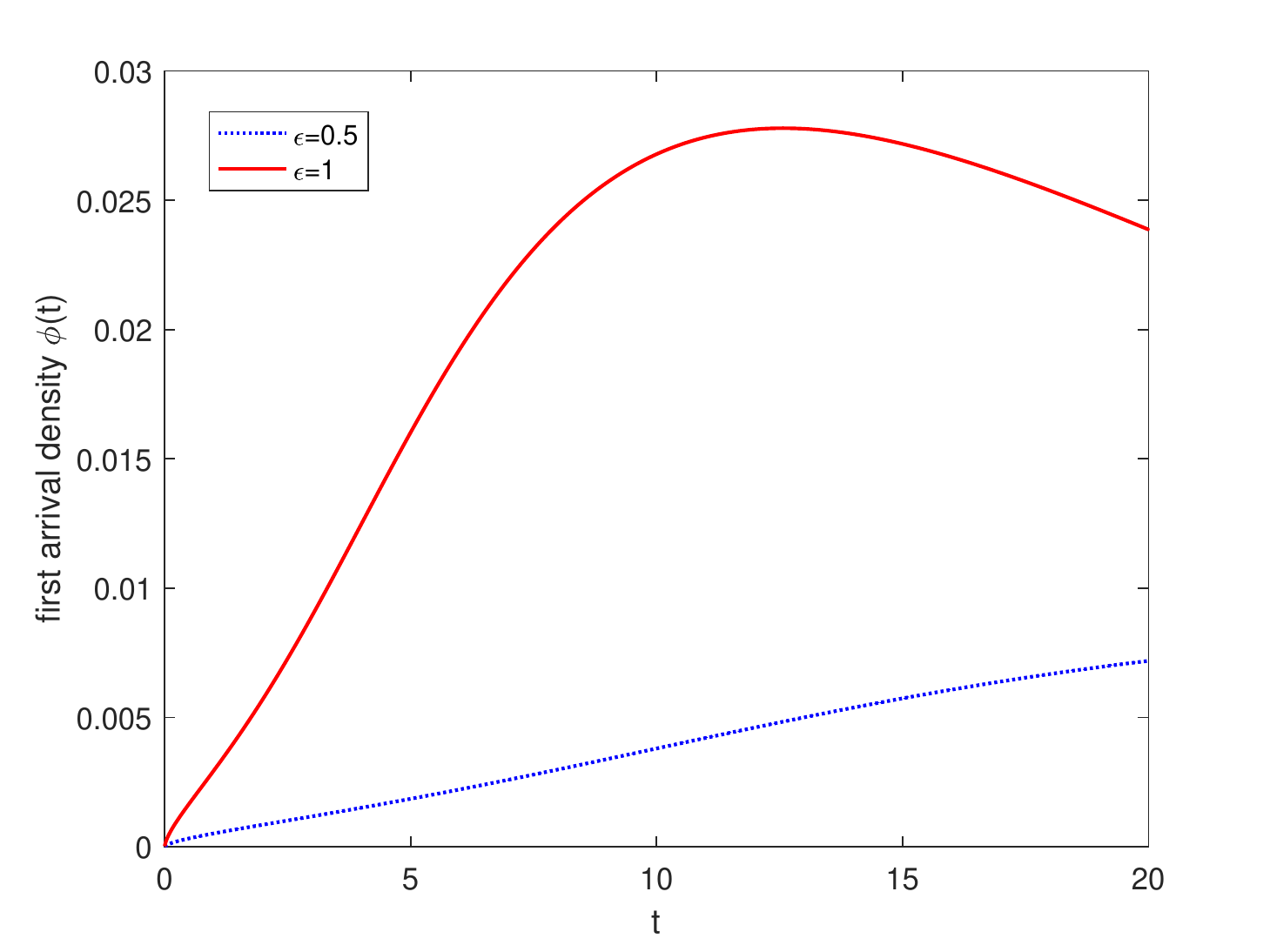}}
\caption{The comparisons of the survival probability and first arrival density with different noise intensities $\epsilon$ for the linear drift case $v(x)=-0.01x$. Here the initial location is $x_0=-0.8$ and the target locates on $x_s=1$. The non-Gaussianity index is $\alpha=1.5$.}
\label{Fig7}
\end{figure}

\indent To explore the influence of the noise intensity on the search mechanism, we plot Figure 6 for the case of small intensity constant  ($\epsilon=0.5$) and compare the survival probability and first arrival density with different noise intensities in Figure 7. Figure 6 indicates that the occurrence of the ``inflection'' on the PDF curve with small noise intensity is much later than that with strong noise as shown in Figure 3. Furthermore, an obvious difference can be seen from Figure 7(a) that the larger noise intensity induces much less survival probability. From Figure 7(b), we infer that larger first arrival density may be obtained with larger noise intensity. The observation implies that a strong noise could lead to a high searching reliability and efficiency, and we can deduce from it that when given the non-Gaussianity index and initial position-target separation, an intensive L\'evy motion can be more helpful for the proteins to find their targets.

\begin{figure}[H]
\centering
\subfigure[]{
\centering
\includegraphics[height = 6cm, width = 10cm]{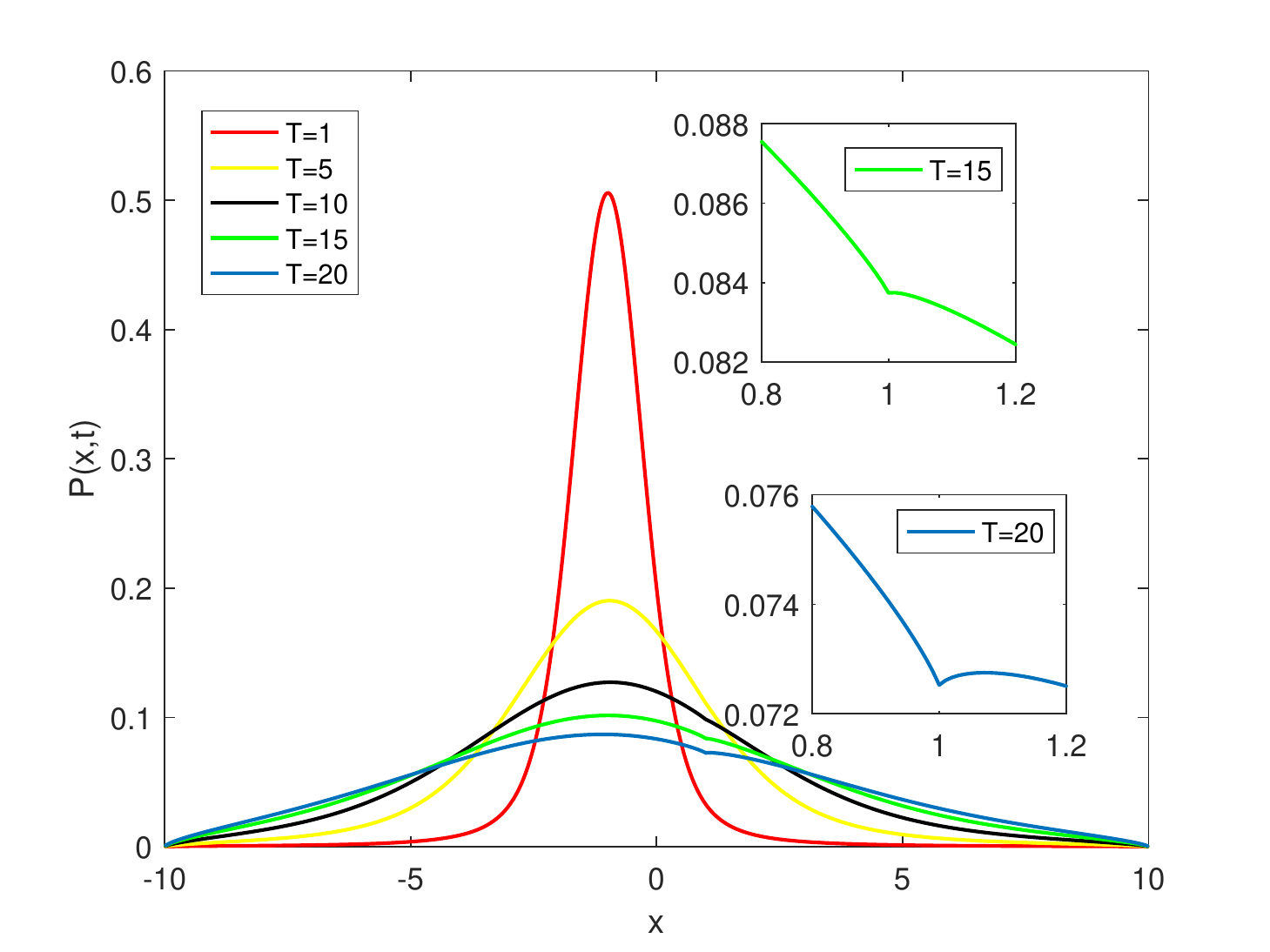}}
\subfigure[]{
\raggedright
\includegraphics[height = 6cm, width = 6cm]{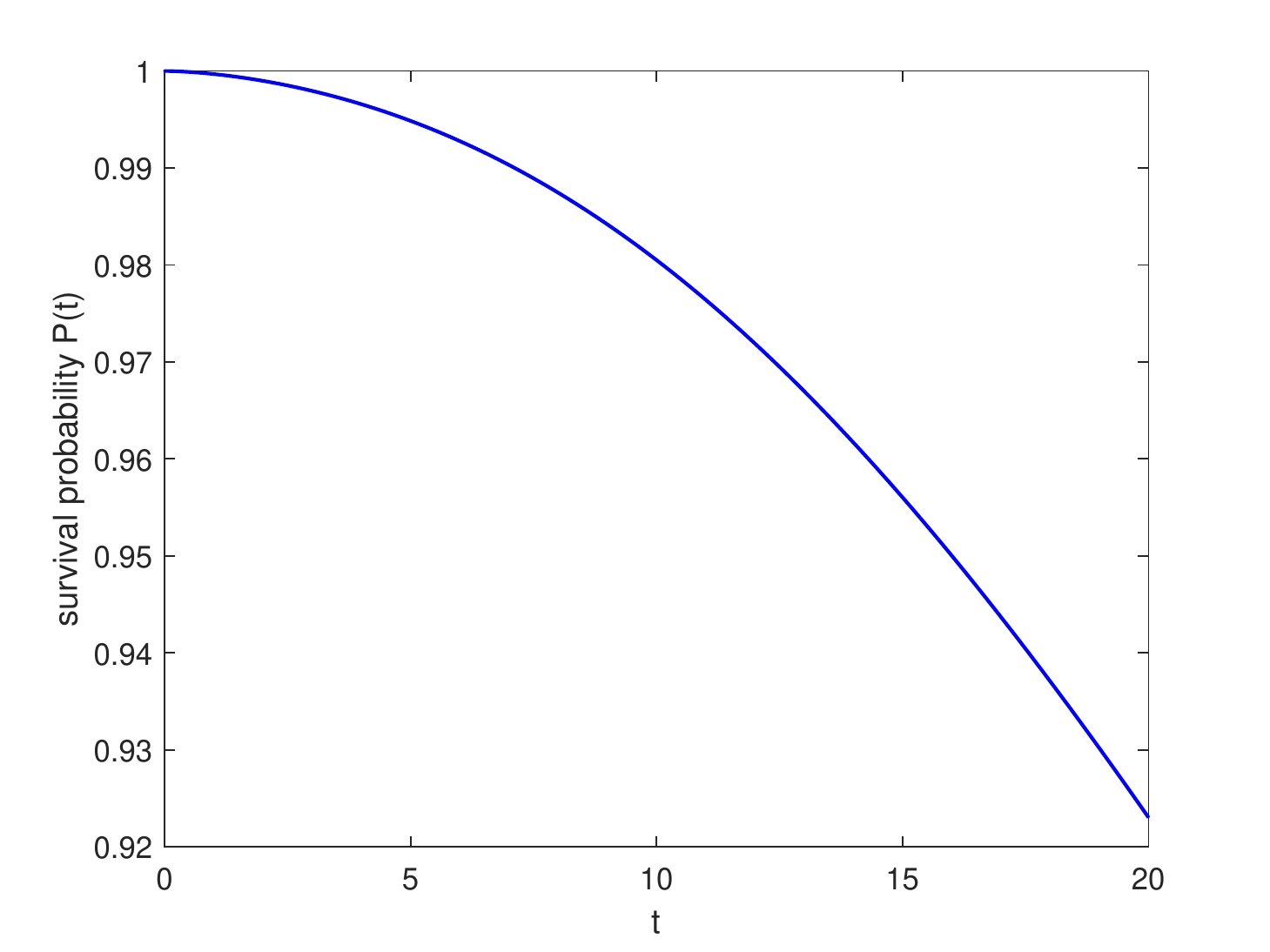}}
\subfigure[]{
\raggedleft
\includegraphics[height = 6cm, width = 6cm]{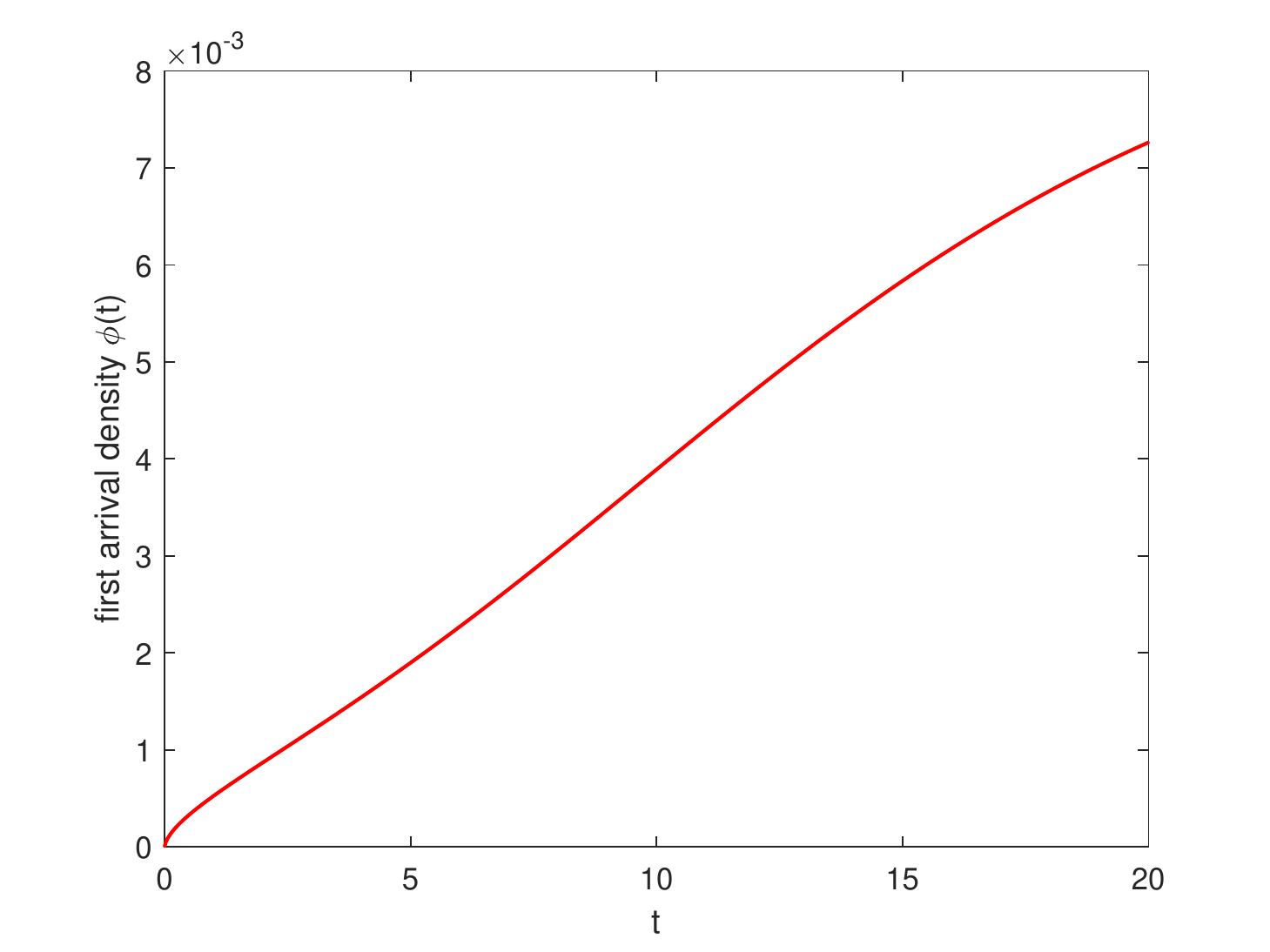}}
\caption{The search strategy in the linear drift case $v(x)=-0.01x$ with the initial location at $x_0=-1$, the target location at $x_s=1$, non-Gaussianity index $\alpha=1.5$, noise intensity $\epsilon=0.5$ and diffusion constant $d=0.1$:
(a) The curves of PDF at different time instants.  (b) The survival probability up to time $t$.  (c) The first arrival density $\phi(t)$.}
\label{Fig8}
\end{figure}

\begin{figure}[H]
\centering
\subfigure[]{
\includegraphics[height = 7.5cm, width = 6cm]{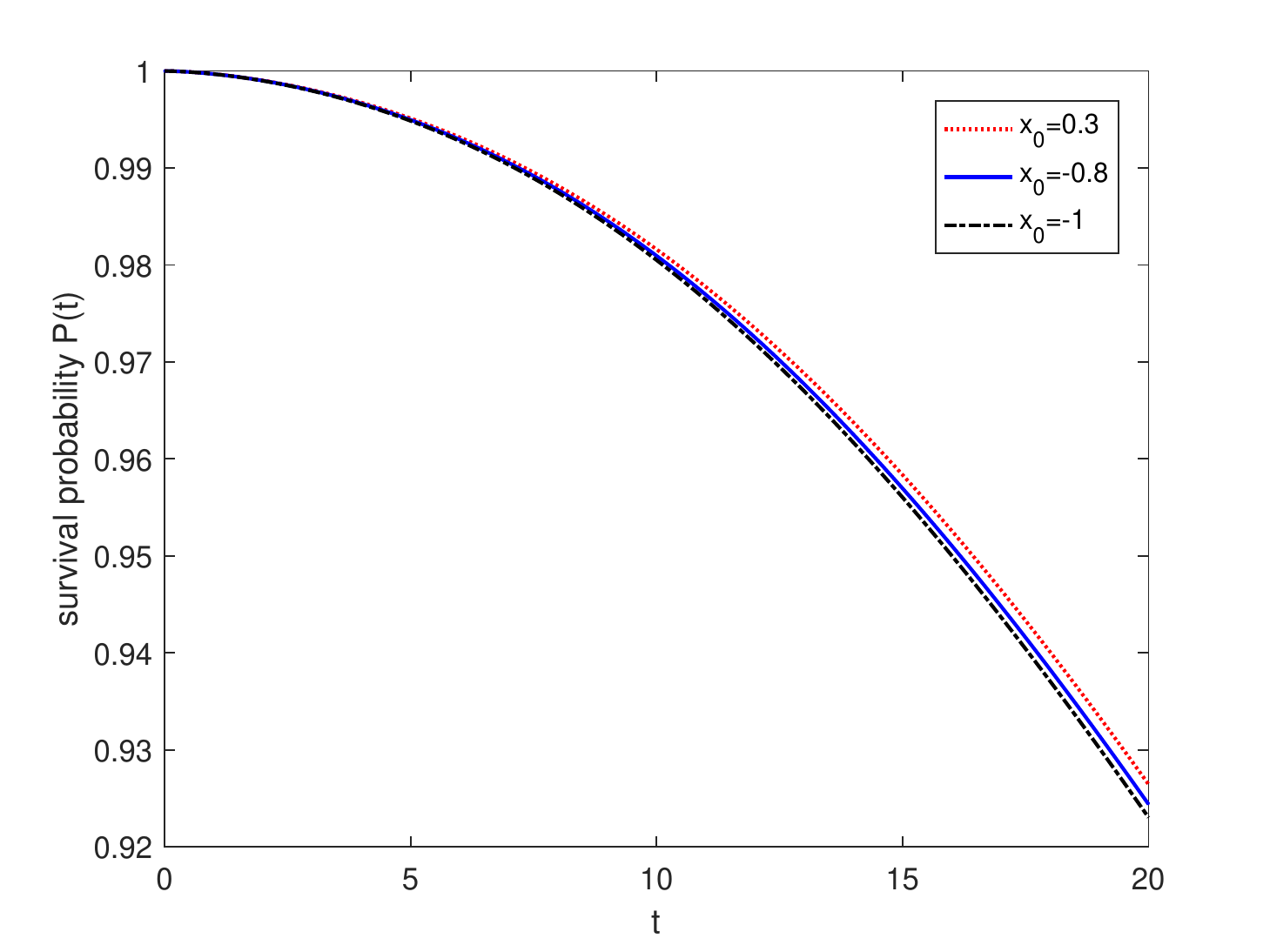}}
\subfigure[]{
\includegraphics[height = 7.5cm, width = 6cm]{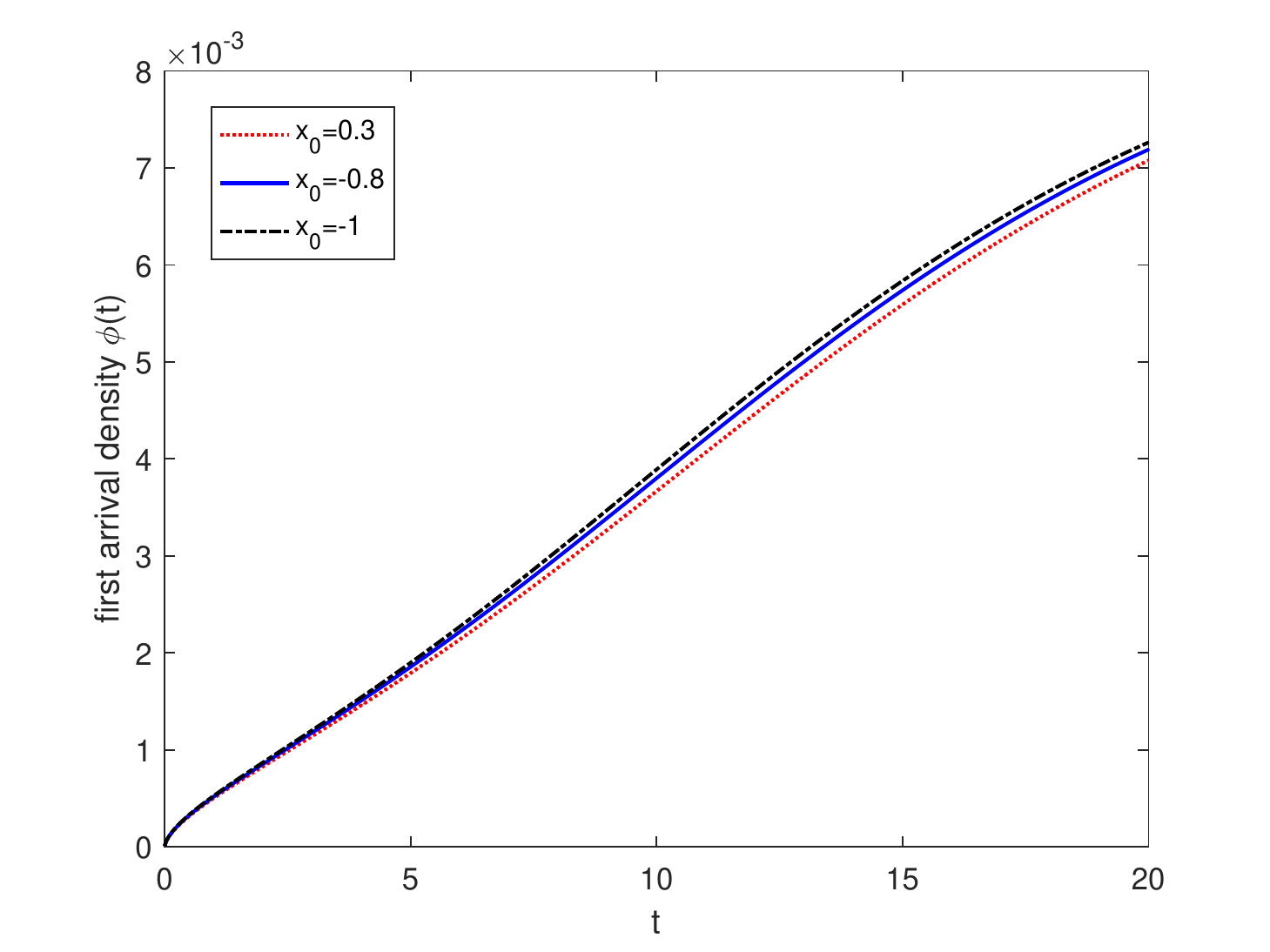}}
\caption{The comparisons of the survival probability and first arrival density with different initial locations by fixing $\alpha=1.5$ and $\epsilon=0.5$ for the linear drift case $v(x)=-0.01x$.}
\label{Fig9}
\end{figure}

\indent For various initial locations of the protein on the DNA, the search efficiencies may also be different. We consider a protein is initially at $x_0=-1$ and the other parameters are kept the same as in Figure 6. Comparisons are shown in Figure 9. It can be seen that when the initial location $x_0=-0.8$ shifts to $x_0=-1$, the survival probability and first arrival density change slightly, which means the initial location does not influence the search reliability remarkably. On the one hand, the subtle difference of the search mechanisms could be attributed to the small distance between the two initial locations. On the other hand, a weak noise intensity could also contribute to the unconspicuous difference. However, we could get some inspiration from the slight disparity. A larger initial position-target separation results in a high search reliability and efficiency when $\alpha=1.5$. This tells us that the LFs search with long-distance jumps dominantly plays the role in avoiding oversampling. Meanwhile, the occurrence of the leap-overs across the target could lead to a low searching efficiency when the initial position is closed to the target.

\subsection{Target search with nonlinear drift: $v(x)=x-x^3$}

\indent In the target searching process, beyond the viscous drag in the nucleus, the macromolecules could act as the blockers to impose  nonlinear resistance on the protein in the searching process. For generality, we consider a double-well function $v(x)=x-x^3$ as the drift to describe the nonlinear force in the search process.

\begin{figure}[H]
\centering
\subfigure[]{
\includegraphics[height = 6cm, width = 10cm]{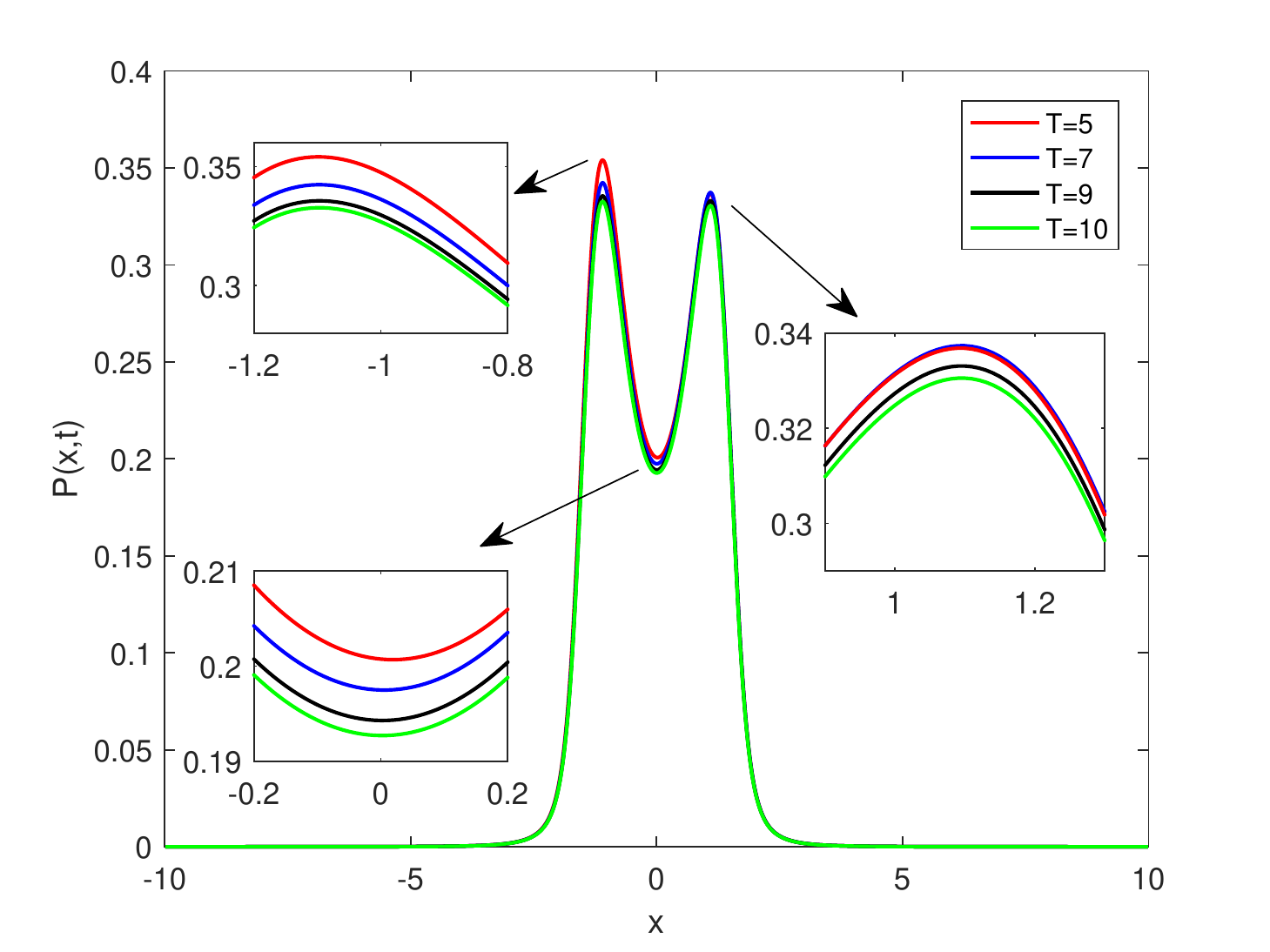}}
\subfigure[]{
\raggedright
\includegraphics[height = 6cm, width = 6cm]{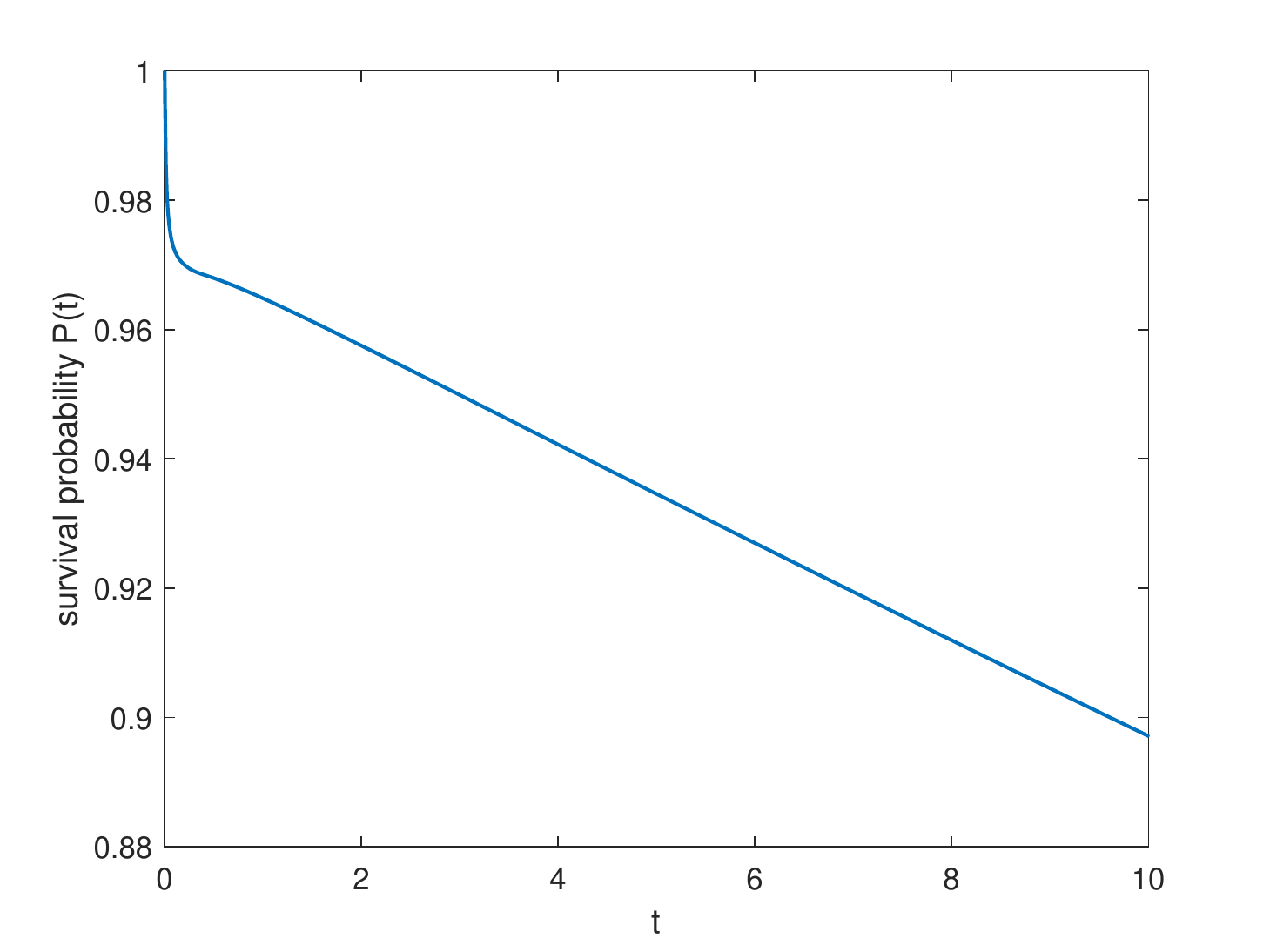}}
\subfigure[]{
\raggedleft
\includegraphics[height = 6cm, width = 6cm]{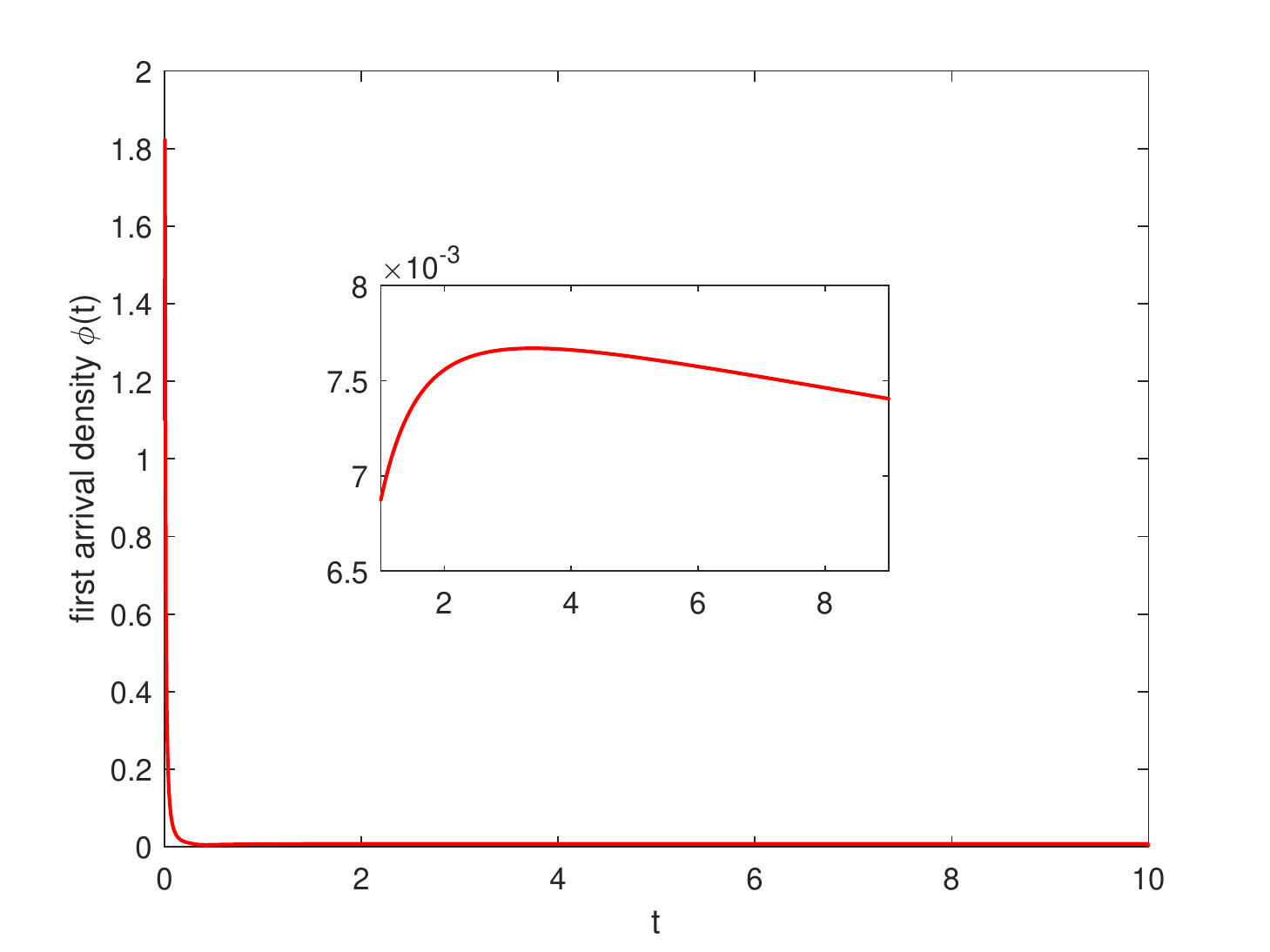}}
\caption{ The search strategy in the nonlinear drift case $v(x)=x-x^3$ with non-Gaussian index $\alpha=1.5$, the initial location at $x_0=-0.8$, the target location at $x_s=1$, noise intensity $\epsilon=1$ and diffusion constant $d=0.1$:
(a) The curves of PDF at different time instants.  (b) The survival probability up to time $t$.  (c) The first arrival density $\phi(t)$. }
\label{Fig10}
\end{figure}

\begin{figure}[H]
\centering
\subfigure[]{
\includegraphics[height = 7.5cm, width = 6cm]{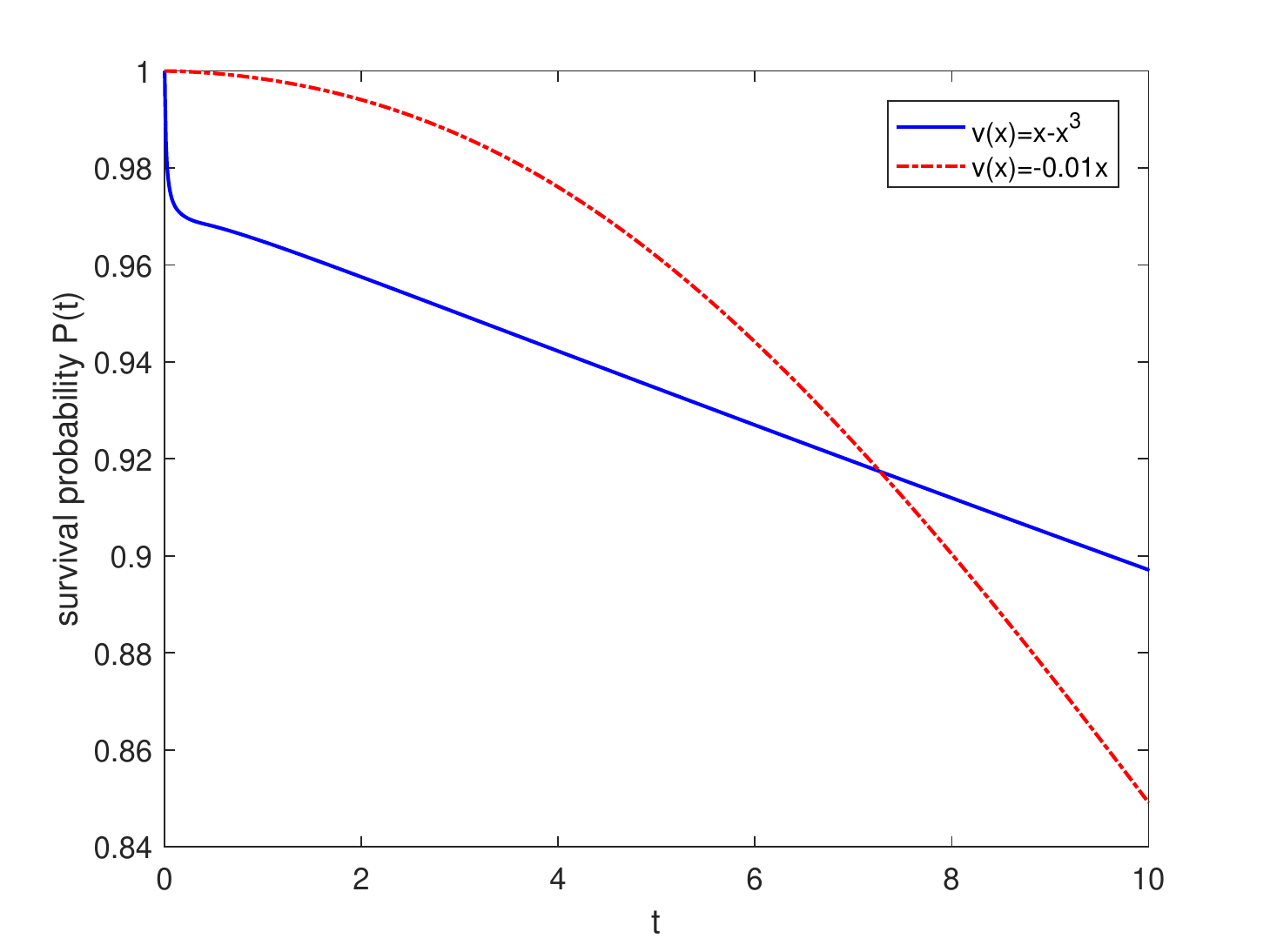}}
\subfigure[]{
\includegraphics[height = 7.5cm, width = 6cm]{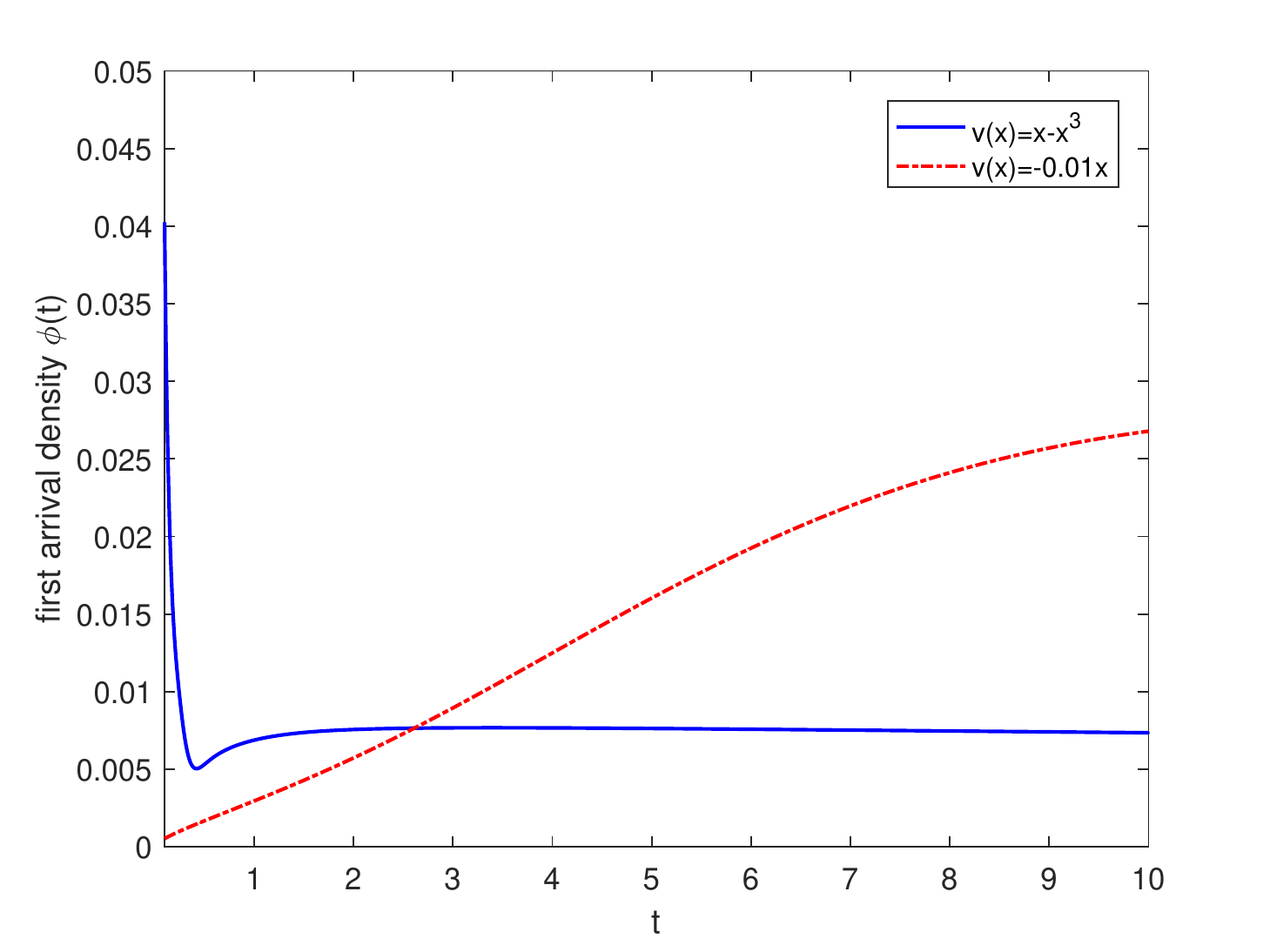}}
\caption{The comparisons of the survival probability and first arrival density with the drift  terms $v(x)=x-x^3$ and $v(x)=-0.01x$ by fixing $\alpha=1.5$ and $\epsilon=1$. We set the initial location at $x_0=-0.8$ and the target located on $x_s=1$.}
\label{Fig11}
\end{figure}

\indent Figure 10 shows the case with nonlinear drift $v(x)=x-x^3$. To see the different influences of the linear and nonlinear drifts on the search strategies intuitively, we put the comparisons on survival probabilities and first arrival densities in Figure 11. It can be found that when the drift is nonlinear, the survival probability experiences a sharp decrease at first. However as time goes on, the survival probability in the nonlinear drift case is larger than that of linear drift. This can be explained by the fact that nonlinear drift could drive the proteins to multiple directions compared with linear drift. Once the initial direction is just towards the target, the protein could find the target site quickly, otherwise the protein would be pushed away from the target location and difficult to reach it. As for the first arrival density, which is defined as the negative derivative of the survival density, a sudden reduction at the very beginning interprets the sharp decrease of the survival density. Then, the evolution of first arrival density tends to constant which corresponds to the approximately linear curve of the survival density.

\begin{figure}[H]
\centering
\subfigure[]{
\includegraphics[height = 7.5cm, width = 6cm]{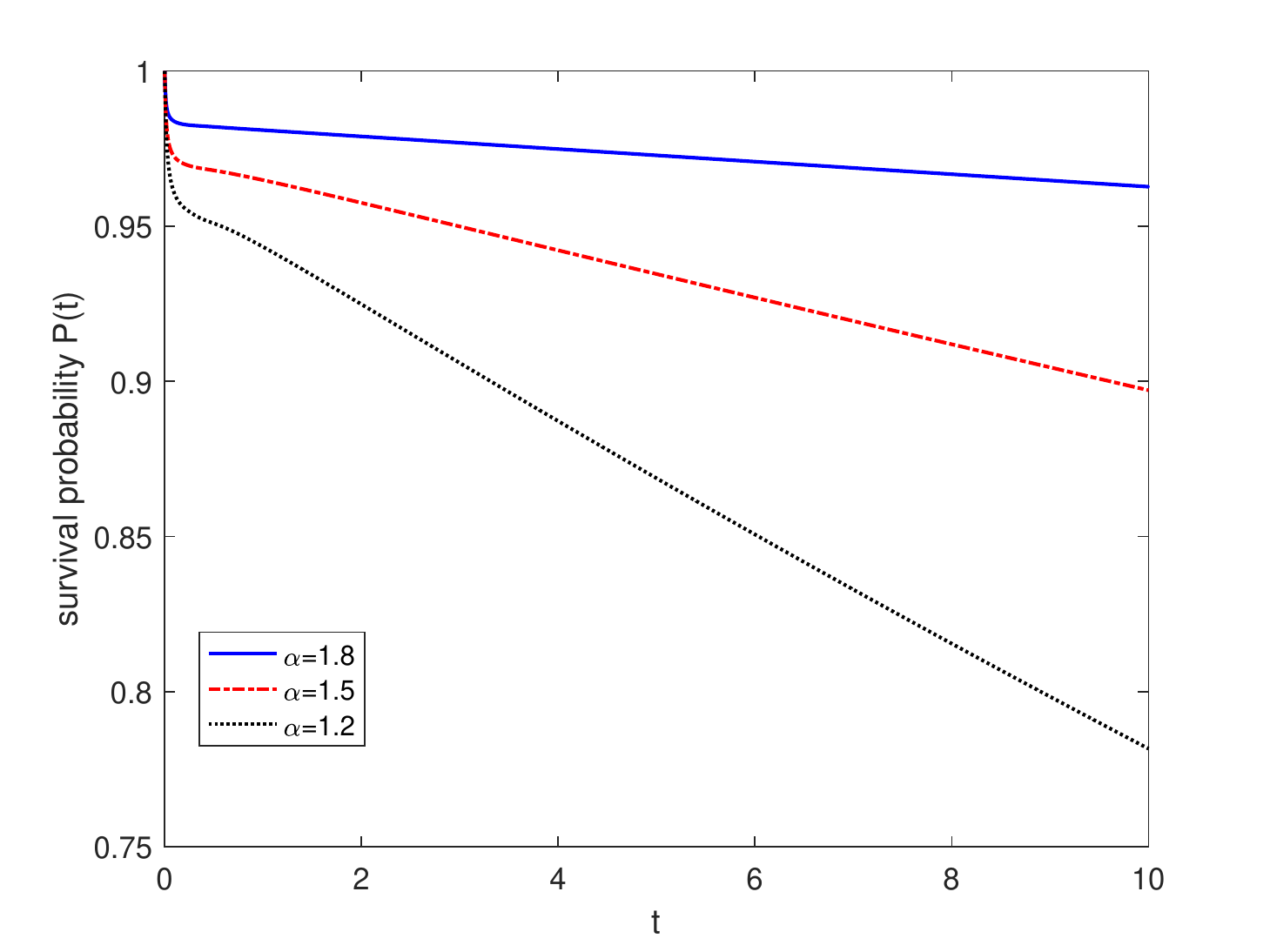}}
\subfigure[]{
\includegraphics[height = 7.5cm, width = 6cm]{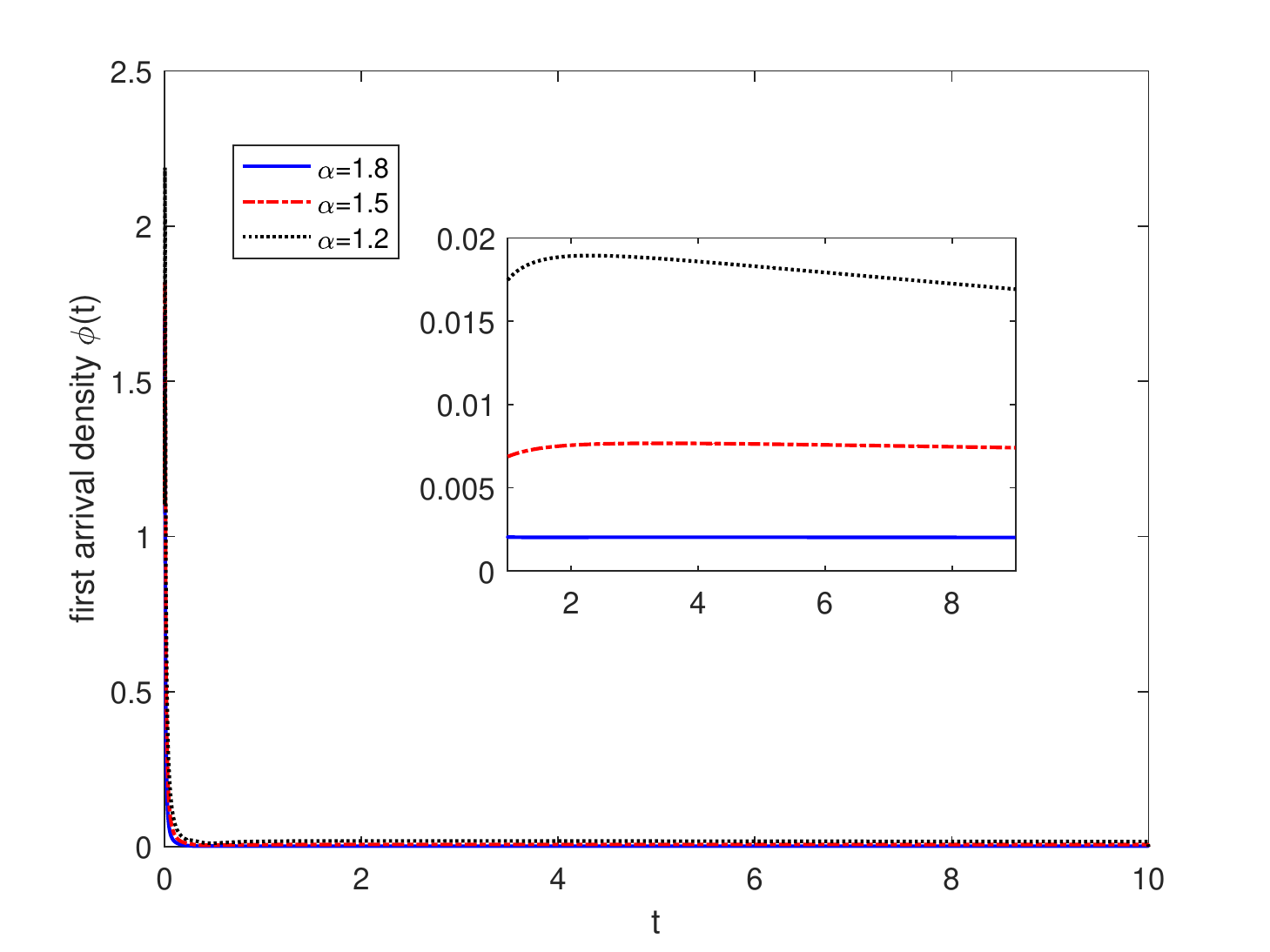}}
\caption{The comparisons of the survival probability and first arrival density in the case of nonlinear drift $v(x)=x-x^3$ with $\alpha=1.2$, $\alpha=1.5$ and $\alpha=1.8$, respectively. We set the initial location at $x_0=-0.8$ and the target located on $x_s=1$. The diffusion intensity is $\epsilon=1$.}
\label{Fig12}
\end{figure}

\indent By comparing the influence of $\alpha$-stable indexes on the searching strategy with nonlinear drift in Figure 12, we notice that a significant disparity occurring among different non-Gaussianity index both for the survival probability and first arrival density. Distinguished with the linear case, the $\alpha$-stable index affects the search reliability and efficiency monotonically, that is to say, the more non-Gaussian the diffusion is, the less survival probability is kept and the larger first arrival density achieves. So under the given initial location and noise intensity, a smaller $\alpha$ will lead to a more reliable and efficient searching process.

\section{Discussion and Conclusion}

In this work, we have investigated the target search mechanisms of the proteins on DNA,  and this  is modelled by a non-local Fokker-Planck equation with $\delta$-sink term. Considering that the proteins move in the nuclear matrix whose density may change with the cell division, we introduce two kinds of drifts into the searching dynamics to describe the external bias. When the nuclear matrix is not crowded, the hindering and hitting interactions among the macromolecules can be ignored and the only drag comes from the solution resistance. In this case, we use a linear function of the position to represent the drift. Otherwise, the double-well function is adopted to depict a nonlinear bias. Here a fast and accurate algorithm for solving the nonlocal Fokker-Planck equation is employed to obtain the PDF of the proteins locating on position $x$ at time $t$. The survival probability and the first arrival density which quantify the search reliability and efficiency are further calculated.

\indent For the searching process with a linear bias, the non-Gaussianity index $\alpha$ displays a non-monotonic effect on the search reliability. Indeed, there is an optimal $\alpha$  at which the search succuss is most likely with the other fixed parameters. Moreover, for the fixed parameters shown in our numerical experiments, the larger the noise intensity, the more possibly or faster that the proteins could find their specific targets. With the small noise intensity, the influences of initial protein-target distance on the search reliability are not evident, but the slight difference also implies  that the proteins initially located far from the target could more likely succeed in searching target with certain $\alpha$ values (for example, $\alpha=1.5$).

\indent Comparing the nonlinear bias case to the linear one, we observe that the nonlinear drift could reduce the search reliability or efficiency with the given parameters, even if it leads to an extremely high possibility for success at initial instant. This   is consistent  with intuitive expectation that the occurrence of bulks will be bound to hinder the search process in the long run, as the direction of particle-moving varies with position in the presence of nonlinear bias. However, if the protein initially moves towards the target without being blocked during very short times, it would quickly find the target. Furthermore, with the nonlinear drift, a smaller non-Gaussianity index $\alpha$ will be more beneficial for the searching process, and this is different from the linear drift case when search reliability varies monotonically with $\alpha$.

\section*{Acknowledgements}

We would like to thank Xu Sun and Xiaofan Li for helpful discussions on numerical simulation. This work was partly supported by the NSF grant 1620449, the NSFC grants 11531006, 11771449, and the program of China Scholarship Council.

\section*{References}


\end{document}